\documentclass[aps,preprintnumbers,eqsecnum,amsmath,amssymb,showpacs,nofootinbib]{revtex4}
\usepackage{graphicx}
\usepackage{bm}
\usepackage{epsfig}
\usepackage{color}
\usepackage{float}

\begin{document}
\title{Rare FCNC radiative leptonic $B_{s,d}\to \gamma l^+l^-$ decays in the Standard Model}
\author{Anastasiia Kozachuk$^{a,b}$, Dmitri Melikhov$^{a,c,d}$, and Nikolai Nikitin$^{a,b,e}$}
\affiliation{
$^a$D.~V.~Skobeltsyn Institute of Nuclear Physics, M.~V.~Lomonosov
Moscow State University, 119991, Moscow, Russia\\
$^b$M.~V.~Lomonosov Moscow State University, Faculty of Physics, 119991, Moscow, Russia\\
$^c$Institute for High Energy Physics, Austrian Academy of Sciences, Nikolsdorfergasse 18, A-1050 Vienna, Austria\\
$^d$Faculty of Physics, University of Vienna, Boltzmanngasse 5, A-1090 Vienna, Austria\\
$^e$A.~I.~Alikhanov Institute for Theoretical and Experimental Physics, 117218 Moscow, Russia}
\date{\today}
\begin{abstract}
We revisit rare radiative leptonic decays $B_{s,d}\to  \gamma e^+e^-$ and $B_{s,d}\to \gamma \mu^+\mu^-$ in the Standard Model and 
provide the updated estimates for various differential distributions (the branching ratios, the forward-backward asymmetry, 
and $R_{\mu/e}$, the ratio of the differential distribution for muons over electrons in the final state). 
The new ingredients of this work compared to the existing theoretical analyses are the following: 
(i) we calculate all $B_d\to\gamma$ and $B_s\to\gamma$ form factors induced by the vector, axial-vector, tensor and 
pseudotensor quark currents within the relativistic dispersion approach based on the constituent quark picture; 
(ii) we perform a detailed analysis of the charm-loop contributions to radiative leptonic decays: we obtain constraints 
imposed by electromagnetic gauge invariance and discuss the existing ambiguities in the charmonia contributions.
\end{abstract}
\maketitle

\section{Introduction}
\label{Sec_introduction}
Rare radiative leptonic $B_{(s,d)}\to \gamma l^+ l^-$ decays are one of the flavor-changing neutral current (FCNC) decays:  
at the quark level, they are induced by $b\to \{s,d\}$ quark transitions, which in the Standard Model (SM) 
are forbidden at tree level. Such transitions occur via penguin and box diagrams containing loops and thus lead to small branching ratios, 
of order $10^{-8}$ -- $10^{-10}$ \cite{Ali:1996vf}. Possible contributions of new particles to the 
loops make these decays particularly sensitive to potential New Physics. 

Several tensions with the SM at the level of 2-4 $\sigma$ have been reported, 
mainly in FCNC $B$-decays (a comprehensive discussion can be found in a recent review \cite{diego2017} and 
\cite{Glashow:2014iga,Guadagnoli:2015nra,Guadagnoli:2016erb}).\footnote{Tensions have 
been reported also in the ratios ${\cal R}_{D^{(*)}}={\cal B}(B\to D^{(*)}\tau\nu)/{\cal B}(B\to D^{(*)}l\nu)$ of the tree-level $B\to D(D^*)$ 
semileptonic decays \cite{Lees:2012xj,Aaij:2015yra,Huschle:2015rga}.} The ratio
\begin{eqnarray}
{\cal R}_K\equiv\frac{{\cal B}(B^+\to K^+\mu^+\mu^-)}{{\cal B}(B^+\to K^+e^+e^-)}=0.745^{+0.090}_{-0.074}(\textrm{stat})\pm 0.036(\textrm{syst})
\end{eqnarray}
in the range $q^2\in[1,6]\,\textrm{GeV}^2$ ($q$ is the momentum of the lepton pair) is 25\% 
lower than the SM prediction at $2.6\sigma$ \cite{Aaij:2014ora, Hiller2a,Hiller2b,Hiller2c}. 
A similar deviation has been recently announced by LHCb in $B^0\to K^{*0}l^+l^-$ decays \cite{lhcb2017web}: 
\begin{eqnarray}
{\cal R}_{K^{*0}}=0.69^{+0.110}_{-0.070}(\textrm{stat})\pm 0.05(\textrm{syst}) \mbox{ for $1.1 < q^2[{\rm GeV}^2] < 6.0.$}
\end{eqnarray}
In an independent measurement, the branching ratio in the region $q^2\in[1,6]\,\textrm{GeV}^2$
\begin{eqnarray}
{\cal B}(B^+\to K^+\mu^+\mu^-)=(1.19\pm 0.03\pm 0.06)\times 10^{-7}
\end{eqnarray}
is 30\% lower than the SM value at $2\sigma$ \cite{Aaij:2014pli,Aaij:2012vr,Hiller1a,Hiller1b,HPQCD};
same was observed for $B^0_s\to\phi\mu^+\mu^-$, in the range  $q^2\in[1,6]\,\textrm{GeV}^2$ the discrepancy for 
the branching ratio is more than $3\sigma$ \cite{Aaij:2015esa}. 
More tensions come from angular analysis of $B\to K^*\mu\mu$ performed by LHCb \cite{Aaij:2015oid} and 
Belle \cite{Abdesselam:2016llu}. Noteworthy, the value of the branching ratio of $B_s\to \mu^+\mu^-$ \cite{CMS:2014xfa} 
is 25\% lower than the SM prediction although only at $1\sigma$.

Obviously, other FCNC $B$-decay modes are good places to search for deviations from the SM. 
The focus of this paper is on rare radiative leptonic $B_{d,s}$-decays.


The $B_{(s,d)}\to \gamma l^+l^-$ decays have been already studied theoretically in a number of publications 
\cite{Geng:2000fs, Dincer:2001hu, Kruger:2002gf, Melikhov:2004mk, Wang:2013rfa, Balakireva:2009kn,dettori2017,gz2017}.
Radiative leptonic $B$-decays have been also extensively discussed in the context of possible lepton flavour violation 
\cite{Glashow:2014iga,Guadagnoli:2015nra,Guadagnoli:2016erb,becirevic,hazard}. 

The previous anylases employed various approximations for the decay amplitudes which excibit a rich structure of nonperturbative QCD effects. 
This work improves the existing analysis in several aspects:  
\begin{itemize}
\item[(i)] 
We calculate all necessary $B_{(s)}\to\gamma$ form factors at timelike momentum transfers using the dispersion 
formulation of the relativistic constituent quark model \cite{ma,mb,melikhov}. This approach proved to be very successful for 
the calculation of numerous meson-to-meson weak transition form factors \cite{ms}; 
in this work we apply this approach to the calculation of the $B\to \gamma$ transition form factors, taking into account rigorous constraints 
on the transition amplitude imposed by electromagnetic gauge invariance. 
\item[(ii)] 
We derive the general gauge-invariance constraints on the charm-loop contributions to the $B\to \gamma l^+l^-$ amplitude. 
We then perform a numerical analysis of charm-loop contributions in $B\to \gamma l^+l^-$ decays, 
including nonfactorizable corrections, making use of the existing results for the $B\to V l^+l^-$ amplitude. 

\item[(iii)] We present a detailed study of a number of observables in $B_{s,d}\to\gamma l^+l^-$ decays (the differential distributions, 
the forward-backward asymmetry, and $R_{\mu/e}$, the ratio of the differential distributions for muons over electrons in the final state, 
which has been recently emphasized in \cite{gz2017} as an interesting observable for radiative leptonic decays).
\end{itemize}
The paper is organized as follows: 
Section \ref{sec_Heff} briefly recalls the effective Hamiltonian for FCNC $b\to s,d$ transitions, and  
Section \ref{sec_amp} describes various contributions to the $B\to \gamma l^+l^-$ amplitude induced by  
$H_{\rm eff}(b\to (s,d)l^+l^-)$. 
In Section \ref{sec_constraints}, we discuss constraints on the $B\to \gamma$ transition amplitude imposed by electromagnetic 
gauge invariance. In Section \ref{sec:cc}, we study in detail contributions to the amplitude of radiative leptonic decay $B\to\gamma l^+l^-$ 
induced by $c$-quark loops, including nonfactorizable effects, and derive rigorous constraints on these contributions imposed by electromagnetic 
gauge invariance. 
Section \ref{sec_kinematics} recalls the differential distributions in radiative leptonic decays. 
Section \ref{sec_ffs} presents the analytic results for the transition form-factors within the dispersion approach 
based on constituent quark picture for all necessary $B\to\gamma$ form factors.  
Section \ref{sec_numerics} contains the numerical predictions for the necessary form factors 
and the observables. Section \ref{conclusions} summarizes our results.

\section{The $b\to d,s$ effective Hamiltonian}
\label{sec_Heff}

A standard theoretical framework for the description of the FCNC $b\to q$ ($q=s,d$) transitions is provided by the Wilson OPE: 
the $b\to q$ effective Hamiltonian describing dynamics at the scale $\mu$, appropriate for $B$-decays, reads 
\cite{Grinstein:1988me,Burasa,Burasb} [we use the sign convention for the effective Hamiltonian 
and the Wilson coefficients adopted in 
\cite{Simulaa,Simulab}]. 
\begin{eqnarray}
\label{Heff}
H_{\rm eff}^{b\to q}=\frac{G_F}{\sqrt{2}}V^*_{tq}V_{tb}\sum_i C_i(\mu) {\cal O}_i^{b\to q}(\mu),  
\end{eqnarray}
$G_F$ is the Fermi constant. The basis operators ${\cal O}_i^{b\to q}(\mu)$ contain only light degrees of freedom   
($u$, $d$, $s$, $c$, and $b$-quarks, leptons, photons and gluons); the heavy degrees of freedom of the 
SM ($W$, $Z$, and $t$-quark) are integrated out and their contributions are encoded in the Wilson coefficients $C_i(\mu)$. 
The light degrees of freedom remain dynamical and 
the corresponding diagrams containing these particles in the loops -- in our case virtual $c$ and $u$ quarks -- 
should be calculated and added to the diagrams generated by the effective Hamiltonian. 

Necessary for the $\bar B_s\to\gamma l^+l^-$ decays of interest are the following terms in (\ref{Heff})\footnote{
Our notations and conventions are: 
$\gamma^5=i\gamma^0\gamma^1\gamma^2\gamma^3$, 
$\sigma_{\mu\nu}=\frac{i}{2}[\gamma_{\mu},\gamma_{\nu}]$, 
$\varepsilon^{0123}=-1$, $\epsilon_{abcd}\equiv
\epsilon_{\alpha\beta\mu\nu}a^\alpha b^\beta c^\mu d^\nu$, 
$e=\sqrt{4\pi\alpha_{\rm em}}$. }(the $\bar B_d\to\gamma l^+l^-$ 
case is obtained with the obvious replacement $s\to d$) \cite{Melikhov:2004mk}:
\begin{eqnarray}
\label{b2qll}
&&H_{\rm eff}^{b\to s l^{+}l^{-}}\, =\, 
{\frac{G_{F}}{\sqrt2}}\, {\frac{\alpha_{\rm em}}{2\pi}}\, 
V_{tb}V^*_{ts}\, 
\left[\,-2im_b\, {\frac{C_{7\gamma}(\mu)}{q^2}}\cdot
\bar s\sigma_{\mu\nu}q^{\nu}\left (1+\gamma_5\right )b
\cdot{\bar l}\gamma^{\mu}l \right.\nonumber\\
&&\left.\qquad\qquad\quad +\, 
C_{9V}(\mu)\cdot\bar s \gamma_{\mu}\left (1\, -\,\gamma_5 \right)   b 
\cdot{\bar l}\gamma^{\mu}l \, +\, 
C_{10A}(\mu)\cdot\bar s   \gamma_{\mu}\left (1\, -\,\gamma_5 \right) b 
\cdot{\bar l}\gamma^{\mu}\gamma_{5}l \right]. 
\end{eqnarray} 
The $C_{7\gamma}$ part in Eq. (\ref{b2qll}) emerges from the diagrams
in Fig.~\ref{fig:A1}a,c with the virtual photon emitted from the penguin 
\begin{equation}
\label{b2qgamma}
H_{\rm eff}^{b\to s\gamma}\, = - \frac{G_{F}}{\sqrt2}\, V_{tb}V^*_{ts}\, 
C_{7\gamma}(\mu)\,\frac{e}{8\pi^2}\, m_b \cdot
\bar s\,\sigma_{\mu\nu}\left (1+\gamma_5\right )b \cdot F^{\mu\nu}.
\end{equation}
Notice that the sign of the $b\to d\gamma$ effective Hamiltonian (\ref{b2qgamma}) correlates with the sign of the electromagnetic vertex. 
For a fermion with the electric charge $Q_qe$, we use in the Feynman diagrams the vertex  
\begin{equation}
iQ_q e \bar q\gamma_\mu q \epsilon^\mu. 
\end{equation}
As already noticed, the light degrees of freedom remain dynamical and their contributions should be taken into account separately. 
The relevant terms in $H_{\rm eff}^{b\to s}$ are those containing four-quark operators: 
\begin{equation}
\label{b2qO12}
H_{\rm eff}^{b\to s\bar cc} = - \frac{G_{F}}{\sqrt2}\, V_{cb}V^*_{cs}\, 
\left\{C_{1}(\mu){\cal O}_1 + C_{2}(\mu){\cal O}_2\right\}
\end{equation}
with 
\begin{eqnarray}
{\cal O}_1=\bar s^j \gamma_\mu(1-\gamma_5)c^i \,\bar c^i \gamma^\mu(1-\gamma_5)b^j, \qquad 
{\cal O}_2=\bar s^i \gamma_\mu(1-\gamma_5)c^i \,\bar c^j \gamma^\mu(1-\gamma_5)b^j, 
\end{eqnarray}
and the similar terms with $c\to u$ ($i,j$ here are color indices). The charm-loop contributions 
generated by operators ${\cal O}_{1,2}$ are discussed in detail in Sect.~\ref{sec:cc}. 

The SM Wilson coefficients at the scale $\mu_0=5$ GeV have the values [corresponding to $C_2(M_W)=-1$]: 
$C_1(\mu_0)=0.241$, 
$C_2(\mu_0)=-1.1$, 
$C_7(\mu_0)=0.312$, 
$C_{9V}(\mu_0)=-4.21$, $C_{10A}(\mu_0)=4.41$ \cite{Burasa,Burasb,Simulaa,Simulab,hidr}. 

\section{Contributions induced by $H_{\rm eff}^{b\to sll}$ and $H_{\rm eff}^{b\to s\gamma}$.\label{sec_amp}} 
In this section, we present the contributions to the $B\to \gamma l^+l^-$ amplitude induced by operators (\ref{b2qll}) and (\ref{b2qgamma}) 
\cite{Melikhov:2004mk}. The $B_s\to \gamma^*$ transition form factors of the corresponding basis operators are defined as \cite{Kruger:2002gf}
\begin{eqnarray}
\label{real}
\label{ffs}
\nonumber
\langle
\gamma^* (k,\,\epsilon)|\bar s \gamma_\mu\gamma_5 b|\bar B_s(p)\rangle 
&=& 
i\, e\,\epsilon^*_{\alpha}\, \left ( g_{\mu\alpha} \, k'k-k'_\alpha k_\mu \right )\,\frac{F_A(k'^2,k^2)}{M_{B_s}}, 
\\
\langle \gamma^*(k,\,\epsilon)|\bar s\gamma_\mu b|\bar B_s(p)\rangle
&=& 
e\,\epsilon^*_{\alpha}\,\epsilon_{\mu\alpha k' k}\frac{F_V(k'^2,k^2)}{M_{B_s}},   
\\
\langle\gamma^*(k,\,\epsilon)|\bar s \sigma_{\mu\nu}\gamma_5 b|\bar B_s(p) 
\rangle\, k'^{\nu}
&=& 
e\,\epsilon^*_{\alpha}\,\left( g_{\mu\alpha}\,k'k- k'_{\alpha}k_{\mu}\right)\, 
F_{TA}(k'^2, k^2), 
\nonumber
\\
\langle
\gamma^*(k,\,\epsilon)|\bar s \sigma_{\mu\nu} b|\bar B_s(p)\rangle\, k'^{\nu}
&=& 
i\, e\,\epsilon^*_{\alpha}\epsilon_{\mu\alpha k' k}F_{TV}(k'^2, k^2).
\nonumber 
\end{eqnarray}
We treat the form factors as functions of two variables, $F_{i}(k'^2,k^2)$: 
here $k'$ is the momentum emitted from the FCNC $b\to q$ vertex, and $k$ is the momentum 
of the photon emitted from the valence quark of the $B$-meson. 
The constraints on the form factors imposed by gauge invariance are discussed in Sect.~\ref{sec_constraints}.

\subsection{\label{sec:A1}
Direct emission of the real photon from valence quarks of the $B$ meson}

\begin{figure}[b!]
\begin{center}
\epsfig{file=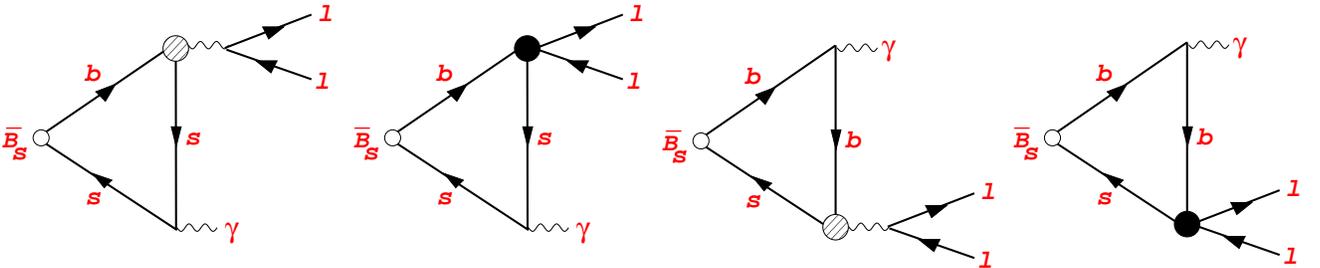,height=3.5cm}
\caption{\label{fig:A1}
Diagrams contributing to $\bar B_s\to \gamma l^+l^-$ discussed in section \protect\ref{sec:A1}.
Dashed circles denote the $b\to s\gamma$ operator $O_{7\gamma}$.    
Solid circles denote the $b\to s l^+l^-$ operators $O_{9V}$ and
$O_{10AV}$.}
\end{center}
\end{figure}

We denote as $A^{(1)}$ the contribution to the $\bar B_s\to \gamma l^+l^-$ amplitude, induced by $H_{\rm eff}^{b\to s l^+l^-}$: 
the real photon is directly emitted from the valence $b$ or $s$ quarks, and the $l^+l^-$ pair is coupled to the FCNC vertex
(see diagrams of Fig. \ref{fig:A1}). 
It corresponds to the momenta $k'=q$, $k=p-q$, $k'^2=q^2$ and $k^2=0$, and thus involves the form factors $F_i(q^2,0)$:
\begin{eqnarray}
\label{A1}
A^{(1)}&=&\langle\gamma (k,\,\epsilon),\,l^+(p_1),\,l^-(p_2)|H_{\rm eff}^{b\to d l^+l^-}|\bar B_s(p) \rangle\, =\,
\frac{G_F}{\sqrt{2}}\, V_{tb}V^*_{tq}\,\frac{\alpha_{\rm em}}{2\pi}\, e\, 
\epsilon^*_{\alpha}
\\
&& \times\left[ 
\epsilon_{\mu\alpha k' k} A_V^{(1)}(q^2)\bar l (p_2)\gamma_{\mu}l (-p_1)
-i\left (g_{\mu\alpha}\, k'k\, -\, k'_{\alpha}k_{\mu}\right)A_A^{(1)}(q^2)\bar l (p_2)\gamma_{\mu}l (-p_1)\right.\nonumber\\
&&\hspace{.2cm}\left.+\epsilon_{\mu\alpha k' k} A_{5V}^{(1)}(q^2)\bar l (p_2)\gamma_{\mu}\gamma_5 l (-p_1)
-i\left (g_{\mu\alpha}\, k'k\, -\, k'_{\alpha}k_{\mu}\right)A_{5A}^{(1)}(q^2)\bar l (p_2)\gamma_{\mu}\gamma_5 l (-p_1)
\right], \quad k'=q,\quad k=p-q.\nonumber
\end{eqnarray}
with 
\begin{eqnarray}
A_{V(A)}^{(1)}(q^2)=\frac{2\, C_{7\gamma}(\mu)}{q^2} m_bF_{TV(TA)}(q^2, 0)+C_{9V}(\mu)\frac{F_{V(A)}(q^2,0)}{M_B},\quad 
A_{5V(5A)}^{(1)}(q^2)=C_{10AV}(\mu)\frac{F_{V(A)}(q^2,0)}{M_B}. 
\end{eqnarray}

\subsection{\label{sec:A2}Direct emission of the virtual photon from valence quarks of the $B$ meson}
\begin{figure}[h!]
\begin{center}
\epsfig{file=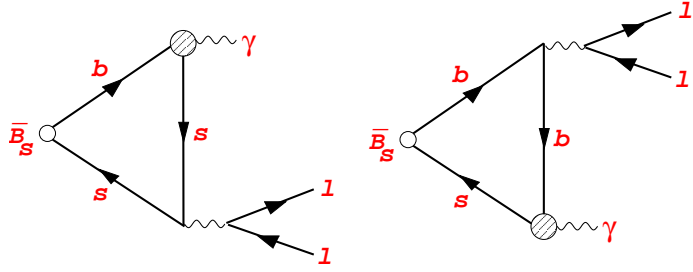,height=3.5cm}
\caption{\label{fig:A2} 
Diagrams describing the $A^{(2)}$ contribution to 
$\bar B_s\to \gamma l^+ l^-$. Dashed circles denote the $b\to s\gamma$ operator $O_{7\gamma}$.}
\end{center}
\end{figure}

Another contribution to the amplitude, $A^{(2)}$, describes the process when the real photon is emitted from the penguin FCNC vertex, 
whereas the virtual photon is emitted from the valence quarks of the $B$-meson (Fig.~\ref{fig:A2}). 
The amplitude $A^{(2)}$ has the same Lorentz structure as the $C_{7\gamma}$ part of $A^{(1)}$ 
where now $k=q$, $k'=p-q$, $k'^2=0$ and $k^2=q^2$. The amplitude thus involves the 
form factors $F_{TA,TV}(0,q^2)$, with $F_{TA}(0,q^2)=F_{TV}(0,q^2)$ (see details in Sect.~\ref{sec_constraints}): 
\begin{eqnarray}
\label{A2}
A^{(2)}&=&\langle\gamma (k',\,\epsilon),\, l^+(p_1),\,l^-(p_2)\left |H_{\rm eff}^{b\to d\gamma} \right|\bar B_s(p) \rangle\,=
\frac{G_F}{\sqrt{2}}\,V_{tb}V^*_{tq} \frac{\alpha_{\rm em}}{2\pi}\, e\,\epsilon^*_{\mu}\bar l (p_2)\gamma_{\alpha} l (-p_1)
\\
&\times&\left[
\varepsilon_{\mu\alpha k'k}A_V^{(2)}(q^2)-i\left(g_{\mu\alpha}k'k - k'_{\alpha}k_{\mu}\right)A_A^{(2)}(q^2)
\right], 
\quad k=q, \quad k'=p-q, \nonumber
\end{eqnarray}
with 
\begin{eqnarray}
A_{V(A)}^{(2)}(q^2)=\frac{2m_b C_{7\gamma}(\mu)}{q^2}F_{TV(TA)}(0, q^2). 
\end{eqnarray}

\subsection{\label{sec:BS}Bremsstrahlung}
\begin{figure}[h!]
\begin{center}
\epsfig{file=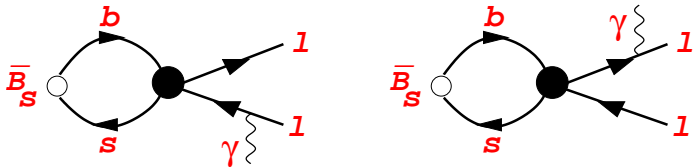,height=2.2cm}
\caption{\label{fig:BS} Diagrams describing photon Bremsstrahlung. Solid circles denote the operator $O_{10A}$.}
\end{center}
\end{figure}

Fig.~\ref{fig:BS} gives diagrams for $A^{\rm Brems}$, the Bremsstrahlung contribution to the $\bar B_s\to \gamma l^+l^-$ amplitude:   
\begin{eqnarray}
\label{bremsstrahlung}
A^{\rm Brems}=-i\, e\,\frac{G_F}{\sqrt{2}}\,\frac{\alpha_{\rm em}}{2\pi}\, V^*_{td}V_{tb}\, 
\frac{f_{B_s}}{M_{B_s}}\, 2\hat m_{l}\, C_{10A}(\mu)\, 
\bar l (p_2)
\left [
\frac{(\gamma\epsilon^*)\,(\gamma p)}{\hat t-\hat m^2_{l}}\, -\, 
\frac{(\gamma p)\,(\gamma\epsilon^*)}{\hat u-\hat m^2_{l}}
\right ]
\gamma_5\, l (-p_1). 
\end{eqnarray}
Let us emphasize (see \cite{Melikhov:2004mk}) that the contribution of the operator $O_{9V}$ to the Bremsstrahlung amplitude vanishes. 

\newpage
\section{Constraints on the transition form factors \label{sec_constraints}}
We now discuss the requirements imposed by the electromagnetic gauge invariance on the 
$\langle  \gamma^*|\bar q O_i b|\bar B_q(p)\rangle$ transition amplitudes induced by the vector, axial-vector, tensor, and pseudotensor weak currents. 
This discussion extends the discussion of \cite{Kruger:2002gf} and includes also the case when the real photon is emitted from the 
FCNC $b\to q$ vertex. The corresponding form factors are functions of two variables, $k'^2$ and $k^2$, 
where $k'$ is the momentum of the weak $b\to q$ current, 
and $k$ is the momentum of the electromagnetic current, $p=k+k'$. 
Gauge invariance provides constraints on some of the form factors describing the transition 
of $B_q$ to the real photon emitted directly from the quark line, i.e. for the form factors at $k^2=0$. 

These form factors fully determine the amplitudes of the FCNC $B$-decays into leptons in the final state. 
For instance, the four-lepton decay of the $B$ meson requires the form factors $f_i(k'^2,k'^2)$ for $0<k^2,k'^2<M_B^2$. 
For the case of the $B\to \gamma l^+l^-$ transition one needs the form factors $f_i(k'^2=q^2,k'^2=0)$ and $f_i(k'^2=0,k'^2=q^2)$, where
$q$ is the momentum of the $l^+l^-$ pair. 

\subsection{Form factors of the vector weak current}
In case of the vector FCNC current, the gauge-invariant amplitude contains one form factor $g(k'^2,k^2)$:
\begin{eqnarray}
\label{vector}
T_{\alpha,\mu}=i\int dx e^{i k x}
\langle 0| T\left\{ j^{\rm e.m.}_\alpha(x), \bar q\gamma_\mu b(0)\right\}|\bar B_q(p)\rangle=
e\,\epsilon_{\mu\alpha k'k} 2g(k'^2,k^2). 
\end{eqnarray}
The amplitude is automatically transverse and is free of the kinematic singularities so no constraints on $g(k'^2,k^2)$ emerge.

\subsection{Form factors of the axial-vector weak current}
For the axial-vector current, the corresponding amplitude has three independent gauge-invariant structures and three form factors, and 
in addition has the contact term which is fully determined by the conservation of the electromagnetic current, $\partial^\mu j_\mu^{e.m.}=0$: 
\begin{eqnarray}
\label{axial-vector}
T^5_{\alpha,\mu}&=&i\int dx e^{i k x}
\langle 0| T\left\{ j^{\rm e.m.}_\alpha(x), \bar q\gamma_\mu\gamma_5 b(0)\right\}|\bar B_q(p)\rangle\nonumber
\\
&=&
i e\,\left(g_{\mu\alpha}-\frac{k_\alpha k_{\mu}}{k^2}\right)f(k'^2,k^2)
+
ie\,\left(k'_{\alpha}-\frac{kk'}{k^2}k_\alpha\right)\bigg[
p_{\mu} a_1(k'^2,k^2) + k_{\mu} a_2(k'^2,k^2)\bigg]
+iQ_{B_q}e\,f_{B_q}\frac{k_\alpha p_\mu}{k^2}.  
\end{eqnarray} 
Here $Q_{\bar B_q}=Q_b-Q_q$ is the electric charge of the $\bar B_q$ meson and $f_{\bar B_q}>0$ is defined according to  
\begin{eqnarray}
\langle 0|\bar q\gamma_\mu\gamma_5 b|\bar B_q(p)\rangle =if_{\bar B_q}p_\mu. 
\end{eqnarray}
The kinematical singularity in the projectors at $k^2=0$ should not be the singularity of the amplitude, and therefore gauge invariance 
yields the following relation between the form factors at $k^2=0$:  
\begin{eqnarray}
\label{constraint1}
\left[f+(k'k)a_2\right]_{k^2=0}=0,\qquad a_1(k'^2,k^2=0)=Q_{\bar B_q} f_{\bar B_q}. 
\end{eqnarray} 
For the neutral $\bar B_{d,s}$ mesons, the contact term is absent and therefore the form factor $a_1$ should vanish at 
$k^2=0$, $a_1(k'^2,k^2=0)=0$. This relation is fulfilled automatically, as the two contributions, corresponding to the  
the photon emission from the valence $b$-quark and from the valence $s,d$-quark cancel each other at $k^2=0$. 

The amplitude of the transition to the real photon is described by a single form factor 
\begin{eqnarray}
\label{axial-vector1}
\langle \gamma(k)|\bar q \gamma_\mu\gamma_5 b|\bar B_q(p)\rangle=
-ie\varepsilon^{\ast\alpha}(k)\left( g_{\mu\alpha} k'k - k'_\alpha k_\mu\right)a_2(k'^2,k^2=0). 
\end{eqnarray}

\subsection{Form factors of the tensor weak current}
The transition amplitudes induced by the tensor weak current can be decomposed in the Lorentz structures transverse 
with respect to $k_\alpha$:  
\begin{eqnarray}
\label{tensor}
T_{\alpha,\mu\nu}&=&i\int dx e^{i k x} \langle 0|\left\{T j_\alpha^{e.m.}(x), \bar q \sigma_{\mu\nu} b(0)\right\}|\bar B_q(p)\rangle \nonumber\\
&=&
ie\,\left(\epsilon_{\mu\nu\alpha p}-\frac{k_\alpha}{k^2}\epsilon_{\mu\nu k p}\right)g_1(k'^2,k^2)
+ie\,\epsilon_{\mu\nu\alpha k}g_2(k'^2,k^2)
+ie\,\left(p_\alpha-\frac{pk}{k^2}k_\alpha\right)
\epsilon_{\mu\nu k' k}g_0(k'^2,k^2),
\end{eqnarray}
The contact terms are absent in this amplitude as well as in the amplitude of the pseudotensor current. 
The kinematic singularity of the projectors at $k^2=0$ should not be the singularity of the amplitude, therefore 
\begin{eqnarray}
\label{constraint_tensor}
\left[g_1-(kp)g_0\right]_{k^2=0}=0. 
\end{eqnarray}
Multiplying (\ref{tensor}) by $k'_\nu$, we obtain the penguin transition amplitude 
\begin{eqnarray}
\label{tensor_penguin}
i\int dx e^{i k x} 
\langle 0|\left\{T j_\alpha^{e.m.}(x), \bar q \sigma_{\mu\nu} k'^\nu b(0)\right\}|\bar B_q(p)\rangle =ie\,\epsilon_{\mu \alpha k p}(g_1+g_2). 
\end{eqnarray}
Notice that the penguin amplitude contains only one combination of the form factors. Nevertheless, 
the requirement of the regularity of the amplitude (\ref{tensor}) yields the constraint (\ref{constraint_tensor}).

\subsection{Form factors of the pseudotensor weak current}
The transition amplitude of the pseudotensor weak current is given in terms of the same form factors as the amplitude (\ref{tensor}), 
and, similar to (\ref{tensor}), contains no contact terms: 
\begin{eqnarray}
\label{pseudotensor}
T^5_{\alpha,\mu\nu}&=&i\int dx e^{i k x} 
\langle 0|\left\{T j_\alpha^{e.m.}(x), \bar q \sigma_{\mu\nu} \gamma_5 b(0)\right\}|\bar B_q(p)\rangle \\
&=& \nonumber
\bigg[
\bigg(g_{\alpha\nu}-\frac{k_\alpha k_{\nu}}{k^2}\bigg)p_\mu-
\bigg(g_{\alpha\mu}-\frac{k_\alpha k_{\mu}}{k^2}\bigg)p_\nu\bigg]e\,g_1
+
(g_{\alpha\nu}k_\mu-g_{\alpha\mu}k_\nu)e\,g_2
+
\bigg(p_{\alpha}-\frac{k\cdot p}{k^2}k_\alpha\bigg)(k_\mu p_\nu-p_\nu k_\mu)e\,g_0. 
\end{eqnarray}
The kinematical singularity in the projectors at $k^2=0$ should cancel in the amplitude, again leading to the constraint 
Eq.~(\ref{constraint_tensor}). 

For the penguin pseudotensor amplitude we then obtain
\begin{eqnarray}
\label{pseudotensor_penguin} 
&&
i\int dx e^{i k x} 
\langle 0|\left\{T j_\alpha^{e.m.}(x), \bar q \sigma_{\mu\nu} \gamma_5 k'^\nu b(0)\right\}|\bar B_q(p)\rangle \nonumber\\
&&=
e\,\left(k'_\alpha k_\mu-g_{\alpha\mu} kk'\right)\left\{g_1+g_2+\frac{k'^2}{kk'}g_1\right\}
+
e\,\left(k'_\alpha-\frac{kk'}{k^2}k_\alpha\right)\left(k_\mu-\frac{kk'}{k'^2}k'_\mu\right)\frac{k'^2}{kk'}\left\{kk' g_0-g_1\right\}. 
\end{eqnarray}
Notice that the contribution of the second Lorentz structure in (\ref{pseudotensor_penguin}) vanishes both for 
$k^2=0$ (because of the constraint Eq.~(\ref{constraint_tensor}): at $k'^2=0$, $kp=kk'$) and for $k'^2=0$. 
However, it does not vanish for both $k^2,k'^2\ne 0$; therefore, the second Lorentz structure contributes to the 
amplitude of the four-lepton decays. 

We can now build the bridge to the form factors which describe the real photon emission by the valence quarks 
defined in Eq.~(\ref{real}): denoting the momentum of the $l^+l^-$ pair as $q$, i.e., setting $k^2=0$ 
and replacing $k'^2\to q^2$, we obtain the form factors in Eq.~(\ref{real}) through the form factors 
$g,a_2,g_2,g_1(k'^2=q^2,k^2=0)$: 
\begin{eqnarray}
\label{rel1}
&&
F_V(q^2,0)=2M_{B} g(q^2,0),\qquad  
F_A(q^2,0)=-M_{B} a_2(q^2,0), \\ 
\label{rel2}
&&
F_{TV}(q^2,0)=-\left[g_2(q^2,0)+g_1(q^2,0)\right],\qquad  
F_{TA}(q^2,0)=-\left[g_2(q^2,0)+\frac{M_B^2+q^2}{M_B^2-q^2}g_1(q^2,0)\right].
\end{eqnarray}
The form factors describing the real photon emission from the penguin, are obtained by setting $k'^2=0$ and replacing 
$k^2\to q^2$ in the form factors $g_{1,2}(k'^2,k^2)$:
\begin{eqnarray}
\label{rel3}
F_{TV}(0,q^2)=F_{TA}(0,q^2)=-[g_2(0,q^2)+g_1(0,q^2)].
\end{eqnarray}
Let us notice that the form factor $g_1(q^2,0)$ should vanish at $q^2=M_B^2$ in order to kill the unphysical pole at $q^2=M_B^2$ in 
the form factor $F_{TA}(q^2,0)$. We shall therefore perform an appropriate subtraction in the spectral representation for $g_1(q^2,0)$ 
to provide this property.

\section{Charm-loop contributions to the amplitude \label{sec:cc}}
Whereas heavy degrees of freedom ($t$, $W$, $Z$) have been integrated out when constructing 
the effective Hamiltonian for $b$-decays, light degrees of freedom, in particular $c$ and $u$ quarks, remain 
dynamical and their contributions in the loops should be taken into account separately. 

We consider in this section the charm-loop contributions to the $B_s\to \gamma l^+l^-$ amplitude, which are related to the following 
matrix element:
\begin{eqnarray}
\label{H}
H_{\mu\alpha}(k',k)=i\int dx e^{i k'x}
\langle 0| T\left\{
\bar c\gamma_\mu c(x), j_\alpha^{\rm e.m.}(0)
\right\}|\bar B_s(p)\rangle, \quad p=k+k'.
\end{eqnarray} 
Here the quark fields are the Heisenberg operators in the SM, i.e. the corresponding $S$-matrix 
includes weak interactions of quarks.

The matrix element (\ref{H}) has the form dictated by the conservation of  
the vector charm-quark and the electromagnetic currents that 
requires $k^\alpha H_{\mu\alpha}(k',k)=0$ and $k'^\mu H_{\mu\alpha}(k',k)=0$ (notice the absence of any contact terms): 
\begin{eqnarray}
\label{H1}
H_{\mu\alpha}(k',k)=
-\frac{G_F}{\sqrt{2}}V_{cb}V^*_{cs}e\left[
\epsilon_{\mu\alpha k'k}H_V
-i\left(g_{\alpha\mu}\,kk'-  k'_\alpha k_\mu\right)H_A
-i\left(k'_\alpha-\frac{kk'}{k^2}k_\alpha\right)\left(k_\mu-\frac{kk'}{k'^2}k'_\mu\right)H_3\right],  
\end{eqnarray} 
with the invariant form factors $H_i$ depending on two variables, $H_i(k'^2,k^2)$. 
The singularities in the projectors at $k^2=0$ and $k'^2=0$ should not be the singularities of the 
amplitude $H_{\mu\alpha}(k',k)$, leading to the constraints 
\begin{eqnarray}
H_3(k'^2=0,k^2)=H_3(k'^2,k^2=0)=0. 
\end{eqnarray}
Let us show that $H_3$ does not contribute to the $B_s\to \gamma l^+l^-$ amplitude:  
to obtain the latter, $H_{\mu\alpha}$ should be multiplied by either $\epsilon^\alpha(k) \bar l\gamma_\mu l$ 
or $\epsilon^\mu(k') \bar l\gamma_\alpha l$. In each case, those terms in the $H_3$-part of $H_{\mu\alpha}$ 
containing $k'_\mu$ or $k'_\alpha$ vanish in the $B_s\to \gamma l^+l^-$ amplitude; the contribution of the ``regular''  structure 
$k_\alpha k_\mu$ also vanishes because the form factor $H_3=0$ if $k^2=0$ or $k'^2=0$.
(The situation is different for the transition into four leptons via two virtual photons, in which case the $H_3$ structure also 
contributes to the $B^0\to l^+l^- l^+ l^-$ transition amplitude). 

\begin{figure}[b]
\begin{center}
\epsfig{file=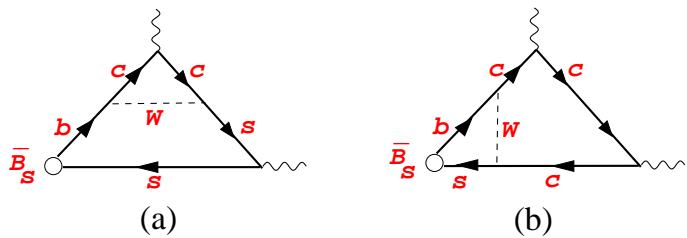,width=9cm}
\caption{\label{Fig:4} 
Lowest-order diagrams describing the contribution of charm to the $B\to \gamma^*\gamma^*$ amplitude:
the charming penguins (a) and the weak-annihilation (b). 
Both diagrams contain CKM factor $V_{cb}V_{cs}^*=-V_{tb}V_{ts}^*$. Figures in this section 
do not display diagrams with the photon emitted from the valence quarks of the $B$-meson.}
\end{center}
\end{figure}

Now, let us consider the matrix element (\ref{H}) at the lowest order in the weak interaction. Fig.~\ref{Fig:4} shows 
the diagrams representing the charm contribution to the $B\to \gamma^*\gamma^*$ amplitude. 
The diagram of Fig.~\ref{Fig:4}a is generated by the $s$-quark part of the electromagnetic current $j^{\rm e.m.}_{\alpha}(0)$ 
[A similar contribution generated by the $b$-quark part of $j^{\rm e.m.}_{\alpha}$ is not shown; it can be easily 
obtained from the $s$-quark part.] The $c$-quark part of $j^{\rm e.m.}_{\alpha}$ generates the diagram of  
Fig.~\ref{Fig:4}b. Integrating out the $W$-boson leads to two different topologies: the charming-penguin 
topology of Fig.~\ref{Fig:4}a and weak-annihilation topology of Fig.~\ref{Fig:4}b.

In addition, the $B\to\gamma l^+l^-$ amplitude receives contributions 
from similar diagrams with the $c$-quark replaced by the $u$-quark.  
The latter, however, contain the CKM factor $V_{ub}V_{us}^*\ll V_{cb}V_{cs}^*$ and are therefore 
strongly suppressed compared to the charm contribution.

\subsection{\label{sec:cp}Charming penguins}
\begin{figure}[t]
\begin{center}
\epsfig{file=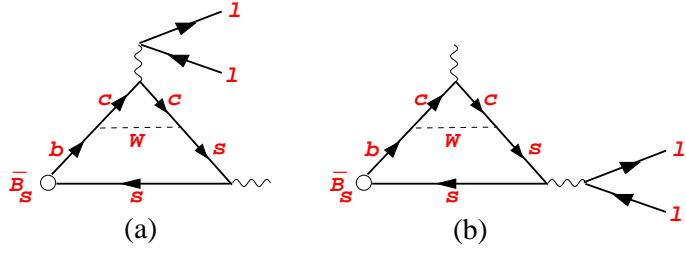,width=9cm}
\caption{\label{Fig:5} 
Two types of the charming-penguin contributions to the $B_s\to \gamma l^+l^-$ amplitude: 
(a) $A^{(1)}_{\bar cc}$, (b) $A^{(2)}_{\bar cc}$. }
\end{center}
\end{figure}

The analytic expression for the charming-penguin diagram of Fig.~\ref{Fig:4}a has the form 
\begin{eqnarray}
\label{Hweak1}
H_{\mu\alpha}(k',k)=\frac{G_F}{\sqrt2}V_{cb}V_{cs}^*
\; i\int dx e^{i k' x}
\langle 0| T\left\{
\bar c \gamma_\mu c(x), 
i \int dy \left[C_1 {\cal O}_1(y)+C_2 {\cal O}_2(y)\right], Q_s e\,\bar s \gamma_\alpha s(0)\right\}|\bar B_s(p)\rangle. 
\end{eqnarray} 
Similar to the diagrams discussed in the previous Section, the diagram of Fig.~\ref{Fig:4}a generates the following two types 
of contributions to the $B_s\to \gamma l^+l^-$ amplitude shown in Fig.~\ref{Fig:5}: the $c$-quark emits the virtual 
photon whereas the $s$-quark emits the real one (a) 
and the $c$-quark emits the real photon whereas the $s$-quark emits the virtual one (b). 

Diagrams of Fig.~\ref{Fig:5} lead to the following contributions to the $B_s\to \gamma l^+l^-$ amplitude: 
\begin{eqnarray}
\label{Acc}
A_{\bar cc}^{(1)}(B_s\to \gamma ll)= H_{\mu\alpha}(k',k)\frac{\bar l\gamma_\mu l}{k'^2}
\varepsilon_\alpha(k)Q_c e^2,\quad k'=q, \quad k=p-q,\nonumber\\
A_{\bar cc}^{(2)}(B_s\to \gamma ll)= H_{\mu\alpha}(k',k)\frac{\bar l\gamma_\alpha l}{k^2}
\varepsilon_\mu(k')Q_ce^2,\quad k=q, \quad k'=p-q. 
\end{eqnarray} 
The Lorentz structure of the amplitudes $A^{(1,2)}_{\bar cc}$ coincides with the Lorentz structure 
of the amplitudes $A^{(1,2)}$, therefore the full charm contribution can be described as additions to the invariant amplitudes 
in Eqs.~(\ref{A1}) and (\ref{A2}):
\begin{eqnarray}
A^{(1)}_{i}(q^2)&\to& \frac{2C_{7\gamma}(\mu)}{q^2} m_b\,F_{Ti}(q^2, 0)+C_{9V}(\mu)\frac{F_{i}(q^2,0)}{M_B}+
\frac{16\pi^2}{3}\frac{H_{i}(q^2,0)}{q^2},\nonumber\\
A^{(2)}_{i}(q^2)&\to& \frac{2C_{7\gamma}(\mu)}{q^2} m_b\,F_{Ti}(0, q^2)+\frac{16\pi^2}{3}\frac{H_{i}(0,q^2)}{q^2},\qquad i=V,A. 
\end{eqnarray} 
The challenging task in the analysis of the charm-loop contributions given by $H_i(q^2,0)$ is the necessity to 
describe a wide range $0 <q^2< M_B^2$, including the region of charmonium resonances. 
Perturbative QCD cannot be applied here and nonperturbative approaches based on hadron degrees of freedom are necessary, 
see discussion in \cite{ali,bbns_duality,hidr,cc_1a,cc_1b,cc_1c,cc_1d,cc_2}. For $H_i(q^2,0)$ one may write dispersion representation in $q^2$ 
with two subtractions, similar to the $B\to K^*l^+l^-$ 
amplitudes $H_{1,2}$ of \cite{hidr}:
\begin{eqnarray}
\label{Hdisp}
H_i(q^2,0)=a_i+ b_i q^2+(q^2)^2\left\{ \sum_{\psi=J/\psi,\psi'} \frac{f_\psi {\cal A}^{i}_{B\psi\gamma}}{m_\psi^3(m_\psi^2-q^2-i m_\psi \Gamma_\psi)}+
h_i(q^2)\right\},\quad i=V,A,   
\end{eqnarray} 
where $a_i$ and $b_i$ are the (unknown) subtraction constants and the functions $h_i(q^2)$ describe the hadron continuum including 
the broad charmonium states lying above the $DD$ threshold. 

The contribution to $A_i^{(1)}$ given by the form factors $H_i(q^2,0)$ may be described as the correction to $C_{9V}$, 
i.e. by the replacement $C_{9V}\to C^{\rm eff}_{9V}(q^2)=C_{9V}+\Delta C_{9V}(q^2)$. Obviously, this correction will be process- and 
Lorentz-structure-dependent: it will in general be different for $B\to P l^+l^-$ decay and for $B\to V l^+l^-$ decay  
and different in the $A_V$ and $A_A$ amplitudes. Describing the nonfactorizable effects as a shift in $C_{9V}$ 
is not particularly convenient in the region of small $q^2$: whereas the full nonfactorizable correction amounts to a few 
percent at small $q^2$ \cite{hidr}, it explodes if expressed as a correction to the coefficient $C_{9V}$. 
Nevertheless, describing the charm-loop effects as corrections to the Wilson coefficients has one very important advantage:
these corrections are obtained as the ratio of the functions $H_i(q^2,0)$ and the appropriate form factors ($B\to K^*$ or $B\to\gamma$). 
It is reasonable to expect that the effects related to the difference between the vector meson and the photon in the final state 
cancel to large extent in the ratios and that the corrections to the Wilson coefficients are approximately equal to each other 
for $B\to \gamma l^+l^-$ and $B\to V l^+l^-$. The accuracy of this approximation is expected to be at the level of 10\%-20\%, the typical 
accuracy of the vector meson dominance. 
 
Similarly, the contributions to $A_i^{(2)}$ given by $H_i(0,q^2)$ may be described as corrections to $C_{7\gamma}$: 
$C_{7\gamma}\to C^{\rm eff}_{7\gamma}(q^2)=C_{7\gamma}+\Delta C_{7\gamma}(q^2)$. 
In principle, this correction is also non-universal and $q^2$-dependent. However, the form factors $H_i(0,q^2)$ and $F_{Ti}(0,q^2)$ 
have similar $q^2$-dependences, as they contain contributions of the same $\bar ss$ hadron resonances in the $q^2$-channel. 
One therefore expects that the correction to the Wilson coefficient $C_{\gamma}$ 
(which is purely nonfactorizable, see the discussion below) may be taken $q^2$-independent. 

In view of these arguments, we will use the results from \cite{hidr} for $\Delta C_9$ and $\Delta C_7$ obtained at 
low $q^2$ for our analysis. 

\subsubsection{Factorizable part of the amplitude $H_{\mu\alpha}$}
Before discussing the full charm-loop corrections to the Wilson coefficients, we present the results for factorizable contributions. 
Taking into account only factorizable gluon exchanges leads to   
\begin{eqnarray}
\label{Hweak2}
H^{\rm fact}_{\mu\alpha}(k',k)&=&\frac{G_F}{\sqrt2}V_{cb}V_{cs}^* \frac{3 C_1+ C_2}{3}\Pi_{\mu\nu }(k')
\left(
i\int dy e^{i k' y}
\langle 0|T\left\{
\bar s \gamma_\nu(1-\gamma_5)b(y),
Q_s e\,\bar s \gamma_\alpha s(0)
\right\}|\bar B_s(p)\rangle\right),  
\end{eqnarray} 
where the expression in brackets is just the amplitude of (\ref{axial-vector}) and 
\begin{eqnarray}
\label{picc1}
\Pi^{cc}_{\mu\nu}(k')=i\int dx e^{i k' x}
\langle 0|
T\{\bar c \gamma_\mu c(x),\bar c \gamma_\nu c(0)\}|0\rangle=
\left(-g_{\mu\nu} k'^2+k'_\mu k'_\nu\right)\Pi^{cc}(k'^2). 
\end{eqnarray}
For the invariant function $\Pi_{cc}(s)$ we may write the spectral representation with one subtraction 
\begin{eqnarray}
\label{picc2}
\Pi_{cc}(k'^2)=\Pi_{cc}(0)+\frac{k'^2}{\pi}\int \frac{{\rm Im}\, \Pi_{cc}(s)}{s(s-k'^2)}ds, 
\end{eqnarray}
At $k'^2\ll 4m_c^2$, $\Pi_{cc}(k'^2)$ can be calculated in perturbative QCD. At leading order in $\alpha_s$, one finds 
\begin{eqnarray}
{\rm Im}\, \Pi_{cc}(s)=\frac{N_c}{12\pi}\frac{2m_c^2+s}{s}\sqrt{1-\frac{4m_c^2}{s}}, \qquad \Pi_{cc}(0)=\frac{9}{16\pi^2}\left\{
-\frac{8}{9}\ln\left(\frac{m_c}{m_b}\right)-\frac{4}{9}\right\}.
\end{eqnarray}
The factorizable contributions to the form factors $H_i(k'^2,k^2)$ are related to $f,a_1,a_2$ as follows
\begin{eqnarray}
\label{Hi}
H_V^{\rm fact}(k'^2,k^2)&=&\frac{3C_1+C_2}{3}k'^2 \Pi_{\bar cc}(k'^2) 2g(k'^2,k^2),\\
H_A^{\rm fact}(k'^2,k^2)&=&\frac{3C_1+C_2}{3}k'^2 \Pi_{\bar cc}(k'^2)\frac{f(k'^2,k^2)}{kk'},\\
H_3^{\rm fact}(k'^2,k^2)&=&\frac{3C_1+C_2}{3}k'^2 \Pi_{\bar cc}(k'^2)\left\{\frac{f}{kk'}+a_1+a_2\right\}.
\end{eqnarray} 
Obviously, $H_{V,A,3}^{\rm fact}(k'^2,k^2)$ vanish for $k'^2=0$.\footnote{The factorizable part of $H_{3}$ vanishes also 
for $k^2=0$ because of the constraints (\ref{constraint1}) on the form factors $f,a_1,a_2$.} 
Therefore, the factorizable $\bar cc$ contribution to the amplitude $A^{(2)}$ and, respectively, to $C_{7\gamma}$, 
vanish; the $\bar cc$ contribution to $A^{(2)}$ comes exclusively from nonfactorizable gluon exchanges. 

The factorizable $\bar cc$ contribution to $A^{(1)}$ can be described as a universal $q^2$-addition to the coefficient $C_{9V}$: 
\begin{eqnarray}
\label{C9eff}
C_{9V}\to C^{\rm eff}_{9V}(q^2)=C_{9V}+\frac{16\pi^2}{9}\left(3 C_1+C_2\right)\Pi_{\bar cc}(q^2). 
\end{eqnarray}

\subsubsection{Adding non-factorizable corrections to the amplitude $H_{\mu\alpha}$}
The functions $H_i(q^2,0)$ may be obtained at $q^2\ll 4m_c^2$ using the method of QCD sum rules. The necessary calculation for 
the $B\to\gamma l^+l^-$ amplitude are not available yet; however, in \cite{hidr} the functions $H_i(q^2,0)$ were calculated for the  
$B\to K^* l^+l^-$ amplitude. As already mentioned above, if the charm-loop effects are described as corrections to the Wilson coefficients, 
the latter are obtained as the ratio of the functions $H_i(q^2,0)$ and the appropriate $B\to V$ form factors ($V=K^*$ or $V=\gamma$). 
There are good reasons to expect that the effects related to the difference in the final states cancel to large extent in these ratios. 
Therefore, we will use the results for $\Delta C_{9V}(q^2)$ and $\Delta C_{7\gamma}(q^2)$ obtained at low $q^2$ from \cite{hidr} 
for our analysis.  

In \cite{hidr} nonfactorizable corrections at low $k^2$ have been calculated using light-cone QCD sum rules. 
The authors noticed that in distinction to the positive-definite factorizable contributions, nonfactorizable corrections 
are not positive-definite and therefore different charmonium resonances may in principle appear with different signs. 
Recall that the absolute values of the amplitudes ${\cal A}^{i}_{B\psi K^*}$ (cf. Eq.~(\ref{Hdisp})) for $\psi$ and $\psi'$ 
are known from the experimental data on $B\to (\psi,\psi') K^*$ decays, but the phases are unknown. 

From our point of view, no conclusion about the relative signs of the resonance contributions may be drawn from the results for 
$\Delta C_{9V}(q^2)$ obtained at $q^2\le 4m_c^2$ where the calculation is trustable. Following \cite{hidr}, we 
describe $h(q^2)$ as an effective heavier resonance of zero width, $h(q^2)=c/(M_R^2-q^2)$. The unknown parameters are now 
the subtraction constants $a,b$, and the parameters of the effective pole $c$ and $M_R^2$. 

\begin{figure}[b]
\begin{center}
\epsfig{file=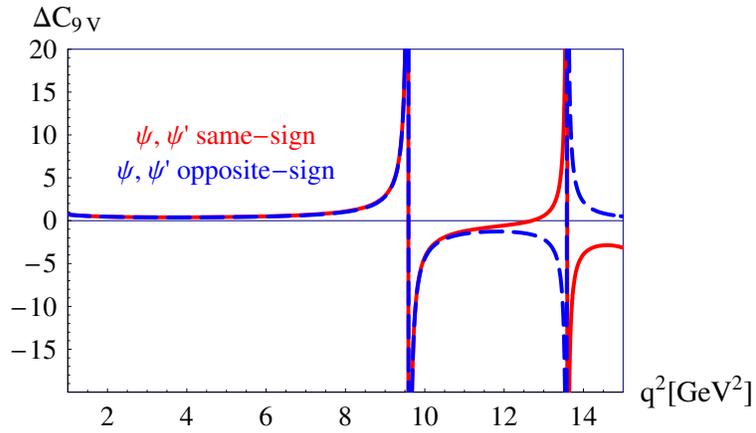,width=10cm}
\caption{\label{Plot:dC9} 
Correction to the Wilson coefficient $C_{9V}$, $\Delta C_{9V}(q^2)$ at $0 < q^2 < 15$ GeV$^2$. 
Red (solid) line correspond to the same-sign positive phases of $\psi $ and $\psi'$; 
Blue (dashed) line correspond to the positive phases of $\psi $ and negative phase of $\psi'$. 
In any case the parameters of subtraction constants $a,b$ and the parameters of the effective pole are determined by a fit to the 
results of \cite{hidr} at $0<q^2<4$ GeV$^2$. In the range $0<q^2<4$ GeV$^2$ both prescriptions for the resonance phases 
reproduce the results from QCD sum rules with better than 1\% accuracy.}
\end{center}
\end{figure}

Fig.~\ref{Plot:dC9} presents two different fits to the results of \cite{hidr} for $\Delta C_{9V}(q^2)$ at $q^2<4$ GeV$^2$
(for our analysis $\Delta C_{9V}^{(B\to K^*,M_{1,2})}$ from \cite{hidr} are relevant; within errors both are equal to each 
other so we take the same $\Delta C_{9V}$ in the vector and the axial-vector amplitudes): 
one fit assumes the standard same positive contributions of $\psi $ and $\psi'$; 
another fit assumes an opposite sign for the $\psi'$ contribution. Obviously, even the knowledge 
of $\Delta C_{9V}(q^2)$ at $q^2 < 4\;{\rm GeV}^2$ with the accuracy of a few percent would 
not allow one to discriminate between the same-phase and the opposite-phase cases. 
Taking into account the expected uncertainty of about 30-50\% of the results from QCD 
sum rules (see Fig.~5 of \cite{hidr}), the question of the relative phases between $\psi$ and $\psi'$ remains fully open.  

We would like to recall that the LHCb collaboration tested the charmonia contributions in the $B\to K l^+l-$ decays but did not manage 
to decide in favour of one or another phase assignments between $\psi$ and $\psi'$. One should take into account, however,  
that in principle the pattern of $\psi$ and $\psi'$ signs may be different in $B\to K l^+l^-$ and in $B\to (K^*,\gamma) l^+l^-$ decays. 

To facilitate calculations at $q^2>4M_D^2$, we take into the contributions of the known broad vector $\psi_n$ ($n=3,\dots,6$) 
resonances (and add a heavier effective pole) to $h(q^2)$ in (\ref{Hdisp}). This may be done by the following addition to $\Delta C_{9V}(q^2)$: 
\begin{eqnarray}
\label{h}
(3C_1+C_2)\frac{3\pi}{\alpha^2_{\rm e.m.}}\sum\limits_{n=3}^6 \left(\frac{q^2}{M_V^2}\right)
\kappa_n \frac{M_V \Gamma(\psi_n\to l^+l^-)}{M_n^2-q^2-i M_n \Gamma_n(q^2)}.
\end{eqnarray}
Here, the factorizable contribution of each $\psi_n$ is multiplied by a fudge factor $\kappa_n$ 
(following the old way of taking into account nonfactorizable corrections \cite{ali}) and a $q^2$-dependent 
$\Gamma_n(q^2)=\frac{(1-q^2/4M_D^2)^{3/2}}{(1-M_n^2/4M_D^2)^{3/2}}\Gamma^{\rm tot}_n$ is used to enable using 
this expression also below the $DD$ threshold. It seems reasonable to take $\kappa_n\sim 1.5-2.0$ for 
all excited vector charmonia: the experimental data for $B\to \psi K^*$ and $B\to \psi' K^*$ lead to 
$\kappa_{J/\psi}=1.6$ and $\kappa_{\psi'}=1.9$. 
The subtraction constants $a,b$ and the parameters of the effective pole are again fixed by requiring 
that $\Delta C_{9V}(q^2)$, corresponding to (\ref{Hdisp}) with the addition (\ref{h}), 
reproduces the sum-rule results at $0<q^2<4$ GeV$^2$. And again, the latter may be easily reproduced with 1\% accuracy.  

To conclude, the results from QCD sum rules available at $q^2\le 4m_c^2$ cannot give a conclusive answer of the relative signs of 
$\psi$ and $\psi'$ resonances (cf. \cite{hidr}). We shall discuss later some observables which are particularly sensitive to the relative 
phases of the $\psi$ and $\psi'$ and could shed light on this issue if measured experimentally.

\subsection{\label{sec:WA}Weak annihilation}
\begin{figure}[h!]
\begin{center}
\epsfig{file=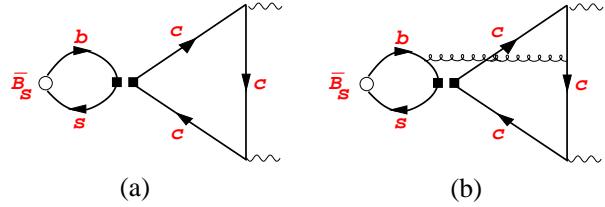,height=2.7cm}
\caption{\label{Fig:6} Weak annihilation diagram: QCD leading-order contribution (a) and nonfactorizable QCD radiative correction (b).}
\end{center}
\end{figure}

Fig.~\ref{Fig:6} shows the typical weak-annihilation (WA) diagrams, which emerge from diagram Fig.~\ref{Fig:4} after 
integrating out the $W$-bosons and taking into account QCD radiative corrections. Diagrams of Fig.~\ref{Fig:6}(a) 
and similar diagrams with gluon exchanges between quark from the same loop lead to factorizable contributions 
$\sim f_{B_s}G_{\gamma \gamma}(k'^2,k^2)$, where the form factor $G_{\gamma \gamma}(k'^2,k^2)$ 
does not depend on the $B_s$-meson structure; the $B_S$-meson contribution is reduced to a single quantity, $f_{B_s}$.  
Diagrams of the type Fig.~\ref{Fig:6}(b) are not reduced to $f_{B_s}$ but contain more complicated quantities describing 
the $B_s$-meson structure. Diagrams with $c$-quarks replaced by $u$-quarks should be included but they are 
CKM suppressed compared to the charm-loop contributions. 

We denote as $A^{\rm WA}$ the corresponding contribution of these diagrams to the $B\to \gamma l^+l^-$ amplitude, and 
we take into account both $c$ and $u$ quarks in the loop. The vertex describing the $\bar bs\to \bar UU$ transition ($U=u,c$) reads
\begin{eqnarray}
\label{heffwa}
H_{\rm eff}^{B_s\to\bar UU} = -\,\frac{G_F}{\sqrt{2}}\, a_1\,V_{Ub}V^*_{Ud}
\,\bar s\gamma_{\mu}(1 -\gamma_5)b 
\,\bar U\gamma_{\mu}(1 -\gamma_5)U, 
\end{eqnarray}
with $a_1\, =\, C_1\, +\, C_2/N_c$, $N_c$ number of colors \cite{stech}. For $N_c\, =\, 3$ 
one finds $a_1\, =\, -0.13$. We now have to take 
\begin{eqnarray}
\langle \gamma l^+l^-|H_{\rm eff}^{B\to\bar UU}|B\rangle.   
\end{eqnarray}
The $\bar UU$ contribution to this amplitude can be written as 
\begin{eqnarray}
A^{WA}(\bar UU)=\frac{G_F}{\sqrt{2}}{V_{Ub}V^*_{Ud}}a_1 2e^3
\epsilon_{\mu \varepsilon^* q k}\frac{G_{\gamma\gamma}(M_B^2,k^2=0,q^2|m^2_U)}{q^2}\,
\bar l^+ \gamma_\mu l^-, 
\end{eqnarray}
where the form factor $G(p^2,k^2,q^2|m^2_U)$ is defined as follows \cite{ms1}
\begin{eqnarray}
\langle \gamma^*(k)\gamma^*(q)|\partial^\nu (\bar U \gamma_\nu \gamma_5 U)|0\rangle
=
e^2\varepsilon^{*\alpha}(k)\varepsilon^{*\beta}(q)
\epsilon_{\alpha\beta kq} G_{\gamma\gamma}(k^2,q^2,p^2|m^2_U).
\end{eqnarray}
For massless $u$-quark in the loop, axial anomaly \cite{abja,abjb} fixes the form factor  
$G_{\gamma\gamma}(p^2,k^2,q^2|0)=-\frac{2N_c (Q_U)^2}{4\pi^2}$. 
For $c$-quark there is an additional $q^2$-dependent contribution given by the amplitude  
$m_c \langle \gamma^*(k)\gamma^*(q)|\bar c \gamma_5 c)|0\rangle \sim
m_c^2/M_B^2$, which contains $\psi$ and
$\psi'$ resonances at $q^2>0$. The latter contribution is numerically
negligible compared to contributions discussed in the previous
sections for all $q^2$ in the reaction of interest. 
Therefore, we have 
\begin{eqnarray}
A^{WA}=-\frac{G_F}{\sqrt{2}}\alpha_{\rm em}e\, a_1
\{V_{ub}V^*_{ud}+V_{cb}V^*_{cd}\}  
\frac{16}{3}
\epsilon_{\mu \varepsilon^* q k}\frac{1}{q^2}\,\bar l^+ \gamma_\mu l^-.  
\end{eqnarray}
The WA contribution is enhanced at small $q^2$, but even here it is suppressed 
by a power of a heavy quark mass compared to the contributions discussed in the previous sections \cite{wa}.

\newpage
\section{\label{sec_kinematics}The $B\to \gamma l^+l^-$ differential distribution}
For convenience, we recall here the results from \cite{Melikhov:2004mk} for the differential distributions. 
The amplitudes discussed in Sections~\ref{sec_amp}(A-C) have the same Lorentz structure, 
whereas the structure of the Bremsstrahlung amplitude (Section \ref{sec:BS}) is different. 
Therefore, it is convenient to write the cross-section as the sum of three contributions: square of the amplitude $A^{1+2+WA}$
which we denote $\Gamma^{(1)}$, square of the amplitude $A^{\rm Brems}$ 
which we denote $\Gamma^{(2)}$, and their mixing, denoted as $\Gamma^{(12)}$ 
(in this Section $F_{V,A}(q^2)$ stands for $F_{A,V}(q^2,0)$):\footnote{In \cite{Melikhov:2004mk}, the  
Bremsstrahlung amplitude Eq.~(2.13) and $\Gamma^{(12)}$ term Eq.~(3.3) had sign errors; these errors are now corrected. 
We are grateful to D.~Guadagnoli for pointing out these sign-errors in \cite{Melikhov:2004mk}.} 
\begin{eqnarray}
\label{Gamma1}
&&\frac{d^2\Gamma^{(1)}}{d\hat s\, d\hat t}\, =\, 
\frac{G^2_F\,\alpha^3_{em}\, M^5_1}{2^{10}\,\pi^4}\, 
\left |V_{tb}\, V^*_{tq} \right |^2
\left [ 
x^2\, B_0\left (\hat s,\,\hat t\right )\, +\,
x\,\,\xi\left (\hat s,\hat t\right )\,\tilde B_1\left (\hat s,\,\hat t\right )
\, +\,  
\xi^2\left (\hat s,\hat t\right )\,\tilde B_2\left (\hat s,\,\hat t\right ) 
\right ], 
\\
\label{BiFi}
&&\qquad\qquad B_0\left (\hat s,\,\hat t\right )\, =\,
    \left (\hat s\, +\, 4\hat m^2_{l} \right )
    \left (F_1\left(\hat s\right )\, +\, F_2\left(\hat s\right )\right)\, -\, 
    8\hat m^2_{l}\,\left |C_{10A}(\mu)\right |^2
    \left (F^2_V\left(q^2\right )\, +\, F^2_A\left(q^2\right )\right), 
    \nonumber \\
&&\qquad\qquad \tilde B_1\left (\hat s,\,\hat t\right )\, =\,
     8\,\left [
                \hat s\, F_V(q^2)\, F_A(q^2)\, 
                Re\left (C^{eff\, *}_{9V}(\mu, q^2)\, C_{10A}(\mu)\right )\, 
        \right .\nonumber\\
&&  \qquad\qquad\qquad\qquad\left . +\,  
                \hat m_b\, F_V(q^2)\, Re\left (C^*_{7\gamma}(\mu)\, 
                \bar F^*_{TA}(q^2)\, C_{10A}(\mu)\right )
              + \hat m_b\, F_A(q^2)\, Re\left (C^*_{7\gamma}(\mu)\, 
                \bar F^*_{TV}(q^2)\, C_{10A}(\mu)\right ) 
        \right ],\nonumber \\
&& \qquad\qquad\tilde B_2\left (\hat s,\,\hat t\right )\, =\,\hat s\, 
    \left (F_1\left(\hat s\right )\, +\, F_2\left(\hat s\right )\right),
   \nonumber\\
&& \qquad\qquad F_1\left (\hat s\right )\, =\, 
   \left (\left |C^{\rm eff}_{9V}(\mu, q^2) \right |^2\, +\,
   \left |C_{10A}(\mu) \right |^2  \right)F^2_V(q^2)
   \, +\,
   \left (\frac{2\hat m_b}{\hat s}\right )^2
   \left |C_{7\gamma}(\mu)\, \bar F_{TV}(q^2)\right |^2\nonumber\\
&& \qquad\qquad \qquad\qquad +\,\frac{4\hat m_b}{\hat s}\, F_V(q^2)\, 
   Re\left (C_{7\gamma}(\mu)\, \bar F_{TV}(q^2)\, C^{eff\, *}_{9V}(\mu, q^2) 
     \right ),\nonumber\\
&& \qquad\qquad F_2\left (\hat s\right )\, =\, 
   \left (\left |C^{\rm eff}_{9V}(q^2, \mu) \right |^2\, +\,
   \left |C_{10A}(\mu)\right |^2  \right)F^2_A(q^2)\, +\,
   \left (\frac{2\hat m_b}{\hat s}\right )^2
   \left |C_{7\gamma}(\mu)\, \bar F_{TA}(q^2)\right |^2\nonumber\\
&& \qquad\qquad\qquad\qquad +\,\frac{4\hat m_b}{\hat s}\, F_A(q^2)\, 
   Re\left (C_{7\gamma}(\mu)\, \bar F_{TA}(q^2)\, C^{eff\, *}_{9V}(\mu, q^2) 
     \right ).\nonumber  
\\
\label{Gamma2}
&&\frac{d^2\Gamma^{(2)}}{d\hat s\, d\hat t} =  
\frac{G^2_F\,\alpha^3_{em}\, M^5_1}{2^{10}\,\pi^4}\, 
\left |V_{tb}\, V^*_{tq} \right |^2\,
\left (\frac{8\, f_{B_q}}{M_B}\right )^2\,\hat m^2_{l}\,
\left |C_{10A}(\mu) \right |^2 
\left [
   \frac{\hat s\, +\, x^2/2}
         {(\hat u\, -\,\hat m^2_{l})(\hat t\, -\,\hat m^2_{l})}\, 
-\,\left (\frac{x\,\hat m_{l}}
        {(\hat u\, -\,\hat m^2_{l})\, (\hat t\, -\,\hat m^2_{l})}
   \right )^2\, 
   \right ]
\\
\label{Gamma12}
&&\frac{d^2\Gamma^{(12)}}{d\hat s\, d\hat t}=
-\frac{G^2_F\,\alpha^3_{em}\, M^5_1}{2^{10}\,\pi^4}\, 
\left |V_{tb}\, V^*_{tq} \right |^2\,\frac{16\, f_{B_q}}{M_B}\, \hat m^2_{l}
\,\frac{x^2}{(\hat u\, -\,\hat m^2_{l})(\hat t\, -\,\hat m^2_{l34})}
 \\ 
&&\qquad\times
\left [
\frac{2\, x\, \hat m_b}{\hat s}\, Re\left (C^*_{10A}(\mu)C_{7\gamma}(\mu)
\bar F_{TV}(q^2, 0)\right )\,
 +\, x\, F_V(q^2)\, Re\left (C^*_{10A}(\mu)C^{\rm eff}_{9V}(\mu, q^2)\right )
  +\,\xi(\hat s,\hat t)\, F_A(q^2)\,\left |C_{10A}(\mu) \right |^2 
  \right ]. \nonumber 
\end{eqnarray}
Here
\begin{eqnarray}
\label{mandelstam}
\hat s\, =\,\frac{\left (p\, -\, k\right )^2}{M_B^2},\quad 
\hat t\, =\,\frac{\left (p\, -\, p_1\right )^2}{M_B^2},\quad 
\hat u\, =\,\frac{\left (p\, -\, p_2\right )^2}{M_B^2}, 
\end{eqnarray} 
with 
$\hat s\, +\,\hat t\, +\,\hat u\, =\, 1\, +\, 2\hat m^2_{l}$, 
$\hat m^2_{l}\, =\,  m^2_{l}/M^2_B$, 
$\hat m_b\, =\, m_b/M_B$ and 
\cite{Kruger:2002gf} 
\begin{eqnarray}
x\, =\, 1\, -\, \hat s,\qquad 
\cos\theta\, =\,\frac{\xi\left (\hat s,\hat t\right )}
          {x\,\sqrt{1\, -\, 4\hat m^2_l/\hat s}},\qquad 
\xi\left (\hat s,\hat t\right )\, =\,\hat u\, -\,\hat t.
\end{eqnarray} 
In the above formulas the complex form factors $\bar F_{TV,TA}$ and defines as follows 
\begin{eqnarray}
\bar F_{TV}(q^2)&=&F_{TV}(q^2,0)+F_{TV}(0,q^2)
-\frac{16}{3}\frac{V_{ub}V^*_{ud}+V_{cb}V^*_{cd}}{V_{tb}V^*_{td}}
\frac{a_1}{C_{7\gamma}}\frac{f_B}{m_b},
\nonumber\\
\bar F_{TA}(q^2)&=&F_{TA}(q^2,0)+F_{TA}(0,q^2).
\end{eqnarray} 
The expressions (\ref{Gamma2}) and (\ref{Gamma12}) contain the
infrared pole which requires a cut in the energy of the emitted
photon. Clearly, the contribution of the pole is proportional to the lepton mass.  


\section{Form factors from the dispersion approach based of the relativistic constituent quark picture\label{sec_ffs}}

So far our discussion was fully general. The problem we face now is to obtain the necessary form factors describing 
the $\bar B\to\gamma$ transition. These form factors are very complicated objects which involve the $\bar B$ mesons in the 
initial state and require a 
treatment with nonperturbative QCD. In the previous section we obtained the constraints on the form factors coming from 
the electromagnetic gauge invariance of the amplitude. There is a number of general constraints on the form factors 
which emerge within large-energy effective theory (LEET) \cite{leet,korch}.  
In our previous analysis we have made use of a simple model for the form factors based on LEET and 
the location of meson singularities in the corresponding channels. The $SU(3)$ breaking effects in the form factors 
have been neglected in that work. 

We now improve our predictions calculating the form factors using the dispersion formulation of the relativistic quark model. 
In this paper we present a model calculation of all necessary form factor within the relativistic dispersion approach based on the 
constituent quark picture. This approach has been formulated in detail in \cite{ma,mb} 
and applied to the weak decays of heavy mesons in \cite{ms}. 

The pseudoscalar meson is described in the dispersion approach by the vertex 
$\bar q_1(k_1)\; \Gamma_5 q(-k_2)\;G(s)/{\sqrt{N_c}}$, where $\Gamma_5=i\gamma_5$ 
with $G(s)=\phi_P(s)(s-M_P^2)$, $s=(k_1+k_2)^2$, $k_1^2=m_1^2$ and $k_2^2=m_2^2$. 
The pseudoscalar-meson wave function $\phi_P$ is normalized according to 
\begin{eqnarray}
\label{norma}
\frac{1}{8\pi^2}\int\limits_{(m_1+m_2)^2}^\infty ds \phi_P^2(s)
\left({s-(m_1-m_2)^2}\right)\frac{\lambda^{1/2}(s,m_1^2,m_2^2)}{s}=1.  
\end{eqnarray}
The decay constant is represented through $\phi_P(s)$ by the spectral integral
\begin{eqnarray}
\label{fP}
f_P=\sqrt{N_c}\int\limits_{(m_1+m_2)^2}^\infty ds \phi_P(s_1)
(m_1+m_2)\frac{\lambda^{1/2}(s,m_1^2,m_2^2)}{8\pi^2s}\frac{s-(m_1-m_2)^2}{s}. 
\end{eqnarray}
Here $\lambda(a,b,c)=(a-b-c)^2-4bc$ is the triangle function. 

The vector meson is described in the dispersion approach by the vertex $\bar q_2(-k_2)\Gamma_\beta q_1(k_1')$, 
with $\Gamma_\beta=\left(A\gamma_\beta+B (k_1'-k_2)_\beta\right)\;G_v(s)/{\sqrt{N_c}}$.  
For the $S$-wave vector meson $A=-1$ and $B=\frac1{\sqrt{s}+m_1+m_2}$. 
Here $s=(k'_1+k_2)^2$, ${k'}_1^2=m_1^2$ and $k_2^2=m_2^2$. The vector-meson wave function 
$\phi_V(s)=G_v(s)/(s-M_V^2)$ is normalized according to 
\begin{eqnarray}
\label{normaV}
\frac{1}{8\pi^2}\int\limits_{(m_1+m_2)^2}^\infty ds \phi_V^2(s)
\left({s-(m_1-m_2)^2}\right)\frac{\lambda^{1/2}(s,m_1^2,m_2^2)}{s}=1.  
\end{eqnarray}
The vector-meson decay constant is represented through $\phi_V(s)$ by the spectral integral \cite{lmss}
\begin{eqnarray}
\label{fV}
f_V=\sqrt{N_c}\int\limits_{(m_1+m_2)^2}^\infty ds \phi_V(s)
\frac{2\sqrt{s}+m_1+m_2}{3}\frac{\lambda^{1/2}(s,m_1^2,m_2^2)}{8\pi^2s}\frac{s-(m_1-m_2)^2}{s}. 
\end{eqnarray}
The photon of virtuality $k^2$ is described by setting $A=\sqrt{N_c}$, $B=0$, $m_2=m_1$, and 
replacing $\phi_v(s)\to \phi_\gamma(s,k^2)$ with 
\begin{eqnarray}
\phi_\gamma(s,k^2)=\frac{1}{s-k^2}. 
\end{eqnarray}
The $P\to \gamma$ form factors $f_i(k'^2,k^2)$ defined in the previous Section 
are obtained as double spectral representations in terms of the relativistic wave function of the $B$-meson 
in the form 
\begin{eqnarray}
\label{doubledisp}
f_i(k'^2,k^2)=\int ds_1 \phi_P(s_1) ds_2\phi_\gamma(s_2,k^2) \Delta_i(s_1,s_2,k'^2|m_2,m_1,m_1). 
\end{eqnarray}
The double spectral representation (\ref{doubledisp}) corresponds to considering the double cut of the 
triangle diagram in variables $p^2$ and $k^2$, treating the variable $k'^2$ as the fixed external current virtuality. 
The double spectral densities $\Delta_i(s_1,s_2,k'^2|m_2,m_1,m_1)$ in variables $p^2$ and $k^2$ 
may be obtained from the known spectral densities of the $P\to V$ transition form factors given by Eqs.~(3.41-3.47) from 
\cite{melikhov} by setting $A=\sqrt{N_c}$ and $B=0$. 

There is, however, also another possibility to obtain the double spectral representation for $f_i(k'^2,k^2)$: one 
can consider the double cut of the triangle diagram in $p^2$ and $k'^2$, at a fixed value of $k^2$. 
\begin{eqnarray}
\label{doubledisp2}
f_i(k'^2,k^2)=\int ds_1 \phi_P(s_1) ds_2\frac{1}{s_2-k'^2}\tilde \Delta_i(s_1,s_2,k^2|m_1,m_1,m_2). 
\end{eqnarray}
The double spectral densities $\tilde \Delta_i(s_1,s_2,k^2|m_1,m_1,m_2)$ differ from 
$\Delta_i(s_1,s_2,k'^2|m_2,m_1,m_1)$, but the form factors calculated from (\ref{doubledisp}) 
and (\ref{doubledisp2}) are of course equal to each other. 
The benefit of using the spectral representations in the form (\ref{doubledisp2}) shows up 
when one considers the transition to the real photon, $k^2=0$: in this case, 
$\tilde \Delta_i(s_1,s_2,k^2\to 0|m_1,m_1,m_2)\to \tilde \rho_i(s_1|m_1,m_2)\delta(s_1-s_2)$, and 
the double spectral representation (\ref{doubledisp2}) is reduced to the single dispersion representation;  
whereas (\ref{doubledisp}) remains a double spectral representation also for $k^2=0$.  

Taking into account this property, we obtain and use the single spectral representations for the form factors 
$F_{A,V,TA,TV}(q^2,0)$, but employ the known double spectral representations in the form (\ref{doubledisp}) 
for $F_{TA,TV}(0,q^2)$. We have checked that for $F_i(q^2,0)$ both representations give the same results.  

\subsection{Form factors $F_i(q^2,0)$ }
We now present single spectral representations for the form factors $F_i(q^2,0)$ [$i=V,A,TV,TA$], corresponding to 
the $\bar B_s$ in the initial state.
\subsubsection{Form factor $F_A(q^2,0)$}
\begin{center}
\begin{figure}[h!]
\mbox{\epsfig{file=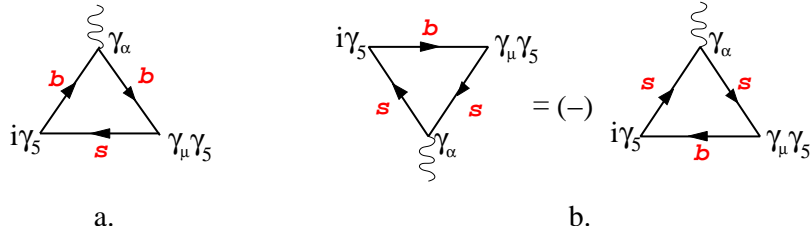,height=3cm}}
\caption{\label{fig:Fa}Diagrams for the form factor $F_A$: (a) $F_A^{(b)}$, (b) $F_A^{(s)}$.} 
\end{figure}
\end{center}

The form factor $F_A$ describing the $\bar B$ in the initial state is given by the diagrams of Fig.~\ref{fig:Fa}. 
Fig.~\ref{fig:Fa}a shows $F_A^{(b)}$, the contribution to the form factor of the process when the $b$ quark emits the photon; 
Fig.~\ref{fig:Fa}b describes the contribution of the process when the quark $d$ emits the photon while $b$ remains a spectator. 
It is convenient to change the direction of the quark line in the loop diagram of Fig.~\ref{fig:Fa}b. 
This is done by performing the charge conjugation of the matrix element and leads to a sign change for the $\gamma_\nu\gamma_5$ vertex. 
Now, both diagrams in Fig.~\ref{fig:Fa}a,b are reduced to the same diagram where quark 1 emits the photon and quark 2 is a spectator: 
setting $m_1=m_b$, $m_2=m_s$ gives $F_A^{(b)}$, while setting $m_1=m_s$, $m_2=m_b$ gives $F_A^{(s)}$ and       
\begin{eqnarray}
F_A=Q_b F_A^{(b)}-Q_s F_A^{(s)}.
\end{eqnarray} 
For the form factor $F_{A}^{(1)}$ a single dispersion integral was obtained in \cite{koz}: 
\begin{eqnarray}
\label{fadisp}
\frac{1}{M_B}F_{A}^{(1)}(q^2,m_1,m_2)&=&\frac{\sqrt{N_c}}{4\pi^2}\int\limits_{(m_1+m_2)^2}^\infty
\frac{ds\;\phi_B(s)}{(s-q^2)}
\left(\rho_+(s,m_1,m_2)+2\frac{m_1-m_2}{s-q^2}\rho_{k_\perp^2}(s,m_1,m_2)\right), 
\end{eqnarray}
where 
\begin{eqnarray}
\label{rhoplus}
\rho_+(q^2,m_1,m_2)&=&(m_2-m_1)\frac{\lambda^{1/2}(q^2,m_1^2,m_2^2)}{s}+m_1\log\left(\frac{q^2+m_1^2-m_2^2+\lambda^{1/2}(q^2,m_1^2,m_2^2)}
{q^2+m_1^2-m_2^2-\lambda^{1/2}(q^2,m_1^2,m_2^2)}\right),
\\
\label{rhokperp2}
\rho_{k_\perp^2}(q^2,m_1,m_2)&=&\frac{q^2+m_1^2-m_2^2}{2q^2}\lambda^{1/2}(q^2,m_1^2,m_2^2)-
m_1^2\log\left(\frac{q^2+m_1^2-m_2^2+\lambda^{1/2}(q^2,m_1^2,m_2^2)}
{q^2+m_1^2-m_2^2-\lambda^{1/2}(q^2,m_1^2,m_2^2)}\right).
\end{eqnarray}
Notice that this expression differs from the analogous expression from \cite{koz}: in the second term we have the factor $1/(s-q^2)$ 
instead of the factor $1/(M_B^2-q^2)$ in Eq. (3.7) of \cite{koz}. 
This corresponds to a slightly different subtraction prescription in the dispersion integral: 
the factor $1/(M_B^2-q^2)$ would lead to the appearance of the unphysical pole at $q^2=M_B^2$. 
For the case of a fixed $q^2=M_V^2$, considered in \cite{koz}, both subtraction prescriptions lead to very close numerical results.

\subsubsection{Form factor $F_V(q^2,0)$} 
\begin{center}
\begin{figure}[ht]
\includegraphics[height=3cm]{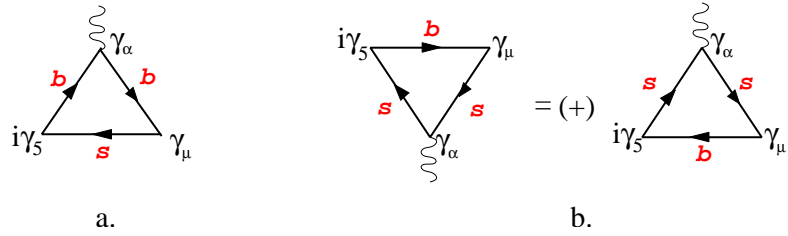}
\caption{\label{fig:Fv}Diagrams for the form factor $F_V$: (a) $F_V^{(b)}$, (b) $F_V^{(s)}$.} 
\end{figure}
\end{center}

The consideration of the form factor $F_{V}$ is very similar to the form factor $F_{A}$. 
$F_V$ is determined by the two diagrams shown in Fig.~\ref{fig:Fv}:  
Fig.~\ref{fig:Fv}a gives $F_V^{(b)}$, the contribution of the process when the $b$ quark interacts with the photon; 
Fig.~\ref{fig:Fv}b 
describes the contribution of the process when the $s$ quark interacts. 
Again, we change the direction of the quark line in the loop diagram of Fig.~\ref{fig:Fv}b by performing the charge 
conjugation of the matrix element. 
For the vector current $\gamma_\nu$ in the vertex the sign does not change (in contrast to the $\gamma_\nu\gamma_5$ case 
considered above). 
Then the contribution of both diagrams in Fig.~\ref{fig:Fv}a,b are given in terms of  
the form factor $F_V^{(1)}(q^2,m_1,m_2)$: Setting $m_1=m_b$, $m_2=m_s$ gives $F_V^{(b)}$ while 
setting $m_1=m_s$, $m_2=m_b$ gives $F_V^{(s)}$, such that   
\begin{eqnarray}
F_V=Q_bF_V^{(b)}+Q_sF_V^{(s)}. 
\end{eqnarray}
The form factor $F_V^{(1)}(q^2,m_1,m_2)$ may be written in the form of a single spectral integral 
\begin{eqnarray}
\frac{1}{M_B}F_V^{(1)}(q^2,m_1,m_2)=-\frac{\sqrt{N_c}}{4\pi^2}\int\limits_{(m_1+m_2)^2}^\infty
\frac{ds\phi_B(s)}{(s-q^2)}\rho_+(s,m_1,m_2). 
\end{eqnarray}

\subsubsection{Form factor $F_{TA}(q^2,0)$}
\label{sec-2.3}
\begin{figure}[H]
\centering
\includegraphics[width=10cm,clip]{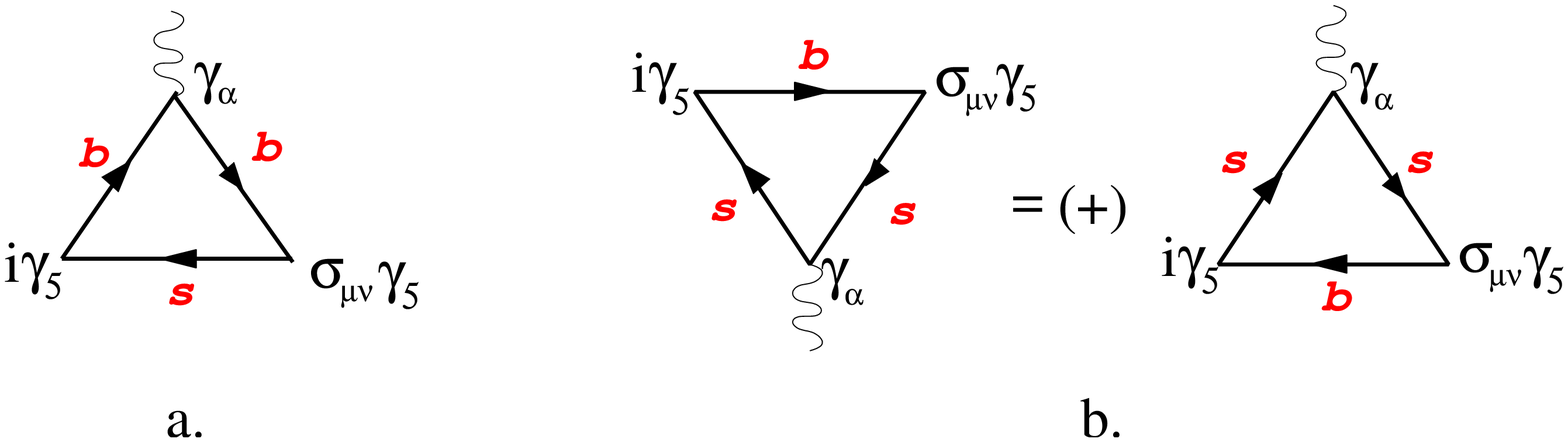}
\caption{Feynman diagrams representing contributions to $F_{TA}$ form factor: (a) $F_{TA}^{(b)}$, (b) $F_{TA}^{(s)}$.}
\label{fig-8}       
\end{figure}

The form factor $F_{TA}$ contains two contributions corresponding to the cases 
when the photon is emitted from $b$ (Fig.~\ref{fig-8}a) and from $d(s)$ (Fig.~\ref{fig-8}b) quark of the B-meson.
Changing the direction of the quark line in the loop diagram of Fig.~\ref{fig-8}b by performing the charge conjugation of the matrix element,
we describe contributions of both diagrams in Fig.~\ref{fig-8} through the same form factor $F_{TA}^{(1)}$:  
setting $m_1=m_b, m_2=m_s$ gives $F^{(b)}_{TA}$, and setting $m_1=m_s, m_2=m_b$ gives $F^{(s)}_{TA}$, such that 
\begin{eqnarray}
\label{FTA}
F_{TA}\,=\,Q_bF^{(b)}_{TA}+Q_sF^{(s)}_{TA}. 
\end{eqnarray}
For the form factor $F^{(1)}_{TA}$ may be written in the form, see (\ref{rel2}),
\begin{eqnarray}			
F_{TA}^{(1)}(q^2,m_1,m_2) &=& -g_2(q^2,m_1,m_2)-\frac{1}{2}(q^2+M_B^2)g_0(q^2,m_1,m_2). 
\end{eqnarray} 
For the form factors $g_{0,2}$ we obtained the following spectral representations 
\begin{eqnarray}
\label{gfunc}
g_0(q^2,m_1,m_2)&=&(M_B^2-q^2)\int\limits_{(m_1+m_2)^2}^\infty ds \frac{\phi_B(s)}{(s-q^2)^2}\rho_{k_\perp^2}(s, m_1, m_2),\nonumber\\
g_2(q^2,m_1,m_2)&=&\int\limits_{(m_1+m_2)^2}^\infty ds \frac{\phi_B(s)}{s-q^2}\rho_{g_2}(s, m_1, m_2),
\end{eqnarray}
with the spectral density  
\begin{eqnarray}
\rho_{g_2}(q^2, m_1, m_2)&=& m_1(m_2-m_1)\log\left(\frac{q^2 + m_1^2 - m_2^2 + \lambda^{1/2}(q^2, m_1, m_2)}
{q^2 + m_1^2 - m_2^2 - \lambda^{1/2}(q^2, m_1, m_2)}\right)+ \lambda^{1/2}(q^2, m_1, m_2),
\end{eqnarray}
and $\rho_{k^2_\perp}$ given by Eq.~(\ref{rhokperp2}). A subtraction has been performed in $g_0$ to provide its vanishing at $q^2=M_B^2$.

\subsubsection{Form factor $F_{TV}(q^2,0)$}
\begin{figure}[H]
\centering
\includegraphics[width=10cm]{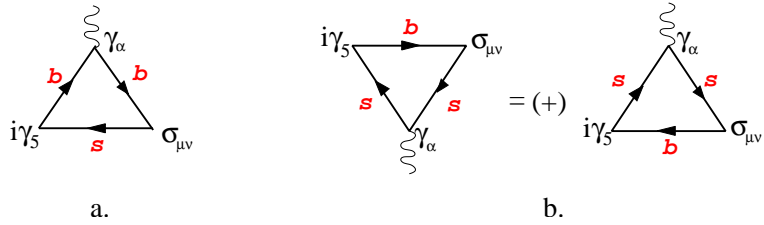}
\caption{Feynman diagrams representing contributions to $F_{TV}$ form factor: (a) $F_{TV}^{(b)}$, (b) $F_{TV}^{(s)}$.}
\label{fig-6}       
\end{figure}

The form factor $F_{TV}$ also contains two contributions corresponding to the cases when the photon is emitted from $b$ or $s$ 
quarks of the $\bar B_s$-meson. These contributions are shown in Fig.~\ref{fig-6}a and Fig.~\ref{fig-6}b.
Changing the direction of the quark line in the loop diagram of Fig.~\ref{fig-6}b by performing the
charge conjugation of the matrix element, we reduce the contribution of each of the diagrams of Fig.~\ref{fig-6}a,b
to the same form factor $F^{(1)}_{TV}$ where quark 1 with mass $m_1$ emits the photon and quark 2 
with mass $m_2$ remains spectator such that  
\begin{eqnarray} 
\label{FTV}
F_{TV}\,=\,Q_bF_{TV}^{(b)}+Q_sF_{TV}^{(s)}.
\end{eqnarray}
The form factor $F^{(1)}_{TV}$ may be written in the form, see (\ref{rel2}),
\begin{eqnarray}				
F_{TV}^{(1)}(q^2,m_1,m_2) &=& - g_2(q^2,m_1,m_2) - \frac{1}{2}(M_B^2-q^2)g_0(q^2,m_1,m_2), 
\end{eqnarray}
with the form factors $g_{0,2}$ given by Eq.~(\ref{gfunc}).


\subsection{Form factors $F_{TA,TV}(0,q^2)$}
The form factor $F_{TA}(0,q^2)=F_{TV}(0,q^2)$, which we denote as $F_T(q^2)$, also has two contributions related to the virtual 
photon emission by the $b$ and the $d(s)$ valence quark. Recall, that the photon emission from the 
heavy valence quark is strongly suppressed compared to the photon emission from the light 
valence quark by a parameter $m_{d,s}/m_b$, where $m_{d,s}$ are the constituent light-quark masses. 

The double spectral representations allow us to calculate the form factors at $q^2<4 m_q^2$, where $m_q$ is the mass of the quark 
which emits the photon. For the case of the photon emission by the $b$-quark, the region $q^2<4 m_b^2$ fully covers 
the decay region 
$0 < q^2 < M_B^2$. The form factor $F^{(b)}(0,q^2)$ in this region is the real function and can be reliably 
calculated within the dispersion approach. 

The situation, however, changes when we consider the process with the virtual photon emission from the light, $d$-or $s$-quark: 
e.g., the physical form factor $F^{(d)}(0,q^2)$ has imaginary part at $q^2>4m_\pi^2$, and contains contribution of light neutral vector-meson 
resonances ($\rho^0$ and $\omega$ for $B$-decays and $\phi$ for $B_s$-decays) in the physical $q^2$-region. 
Obviously, our calculation based on the quark degrees of freedom is trustable (far) from the hadron thresholds 
and should not be applied in the region of hadron resonances. 

To obtain the form factor $F_T(0,q^2)$ at $0< q^2<M_B^2$ we therefore proceed as follows:\footnote{Another possibility would be 
to calculate the form factor $F^{(b)}(0,q^2)$ using the dispersion approach and to apply Vector-Meson Dominance (VMD) to 
$F^{(d,s)}(0,q^2)$. 
However, $F^{(b)}(0,q^2)$ is a relatively flat function at $0<q^2<M_B^2$ (it has pole at $q^2=M_\Upsilon^2$), 
and provides numerically small contribution at the level of about 10\% compared to $F^{(d,s)}(0,q^2)$. 
We therefore find eligible to apply the VMD approximation to the full form factor $F^{(d,s)}(0,q^2)$.}
we calculate the form factors $F^{(d,s)}(0,q^2)$ using the gauge-invariant version \cite{Melikhov:2003hs} 
of the vector meson dominance \cite{Sakurai:1960ju,GellMann:1961tg,Gounaris:1968mw}
\begin{eqnarray}
\label{vmd}
F_{TV,TA}(0, q^2) = F_{TV,TA}(0, 0)\, -\,\sum_V\,2\,f_V^{\rm e.m.} g^{B\to V}_+(0)\frac{q^2/M_V}{q^2\, -\, M^2_V\, +\, iM_V\Gamma_V},
\end{eqnarray}
where $M_V$ and $\Gamma_V$ are the mass and the width 
of the vector meson resonance, 
$g^{B\to V}_+(0)$ are the $B\to V$ transition form factors, 
defined according to the relations 
\begin{eqnarray*}
\langle V(q, \varepsilon)|\bar d\sigma_{\mu\nu} b|B(p)\rangle
\, =\, i\varepsilon^{*\alpha}\,\epsilon_{\mu\nu\beta\gamma}
\left[ 
g^{B\to V}_+(k^2)g_{\alpha\beta}(p+q)^{\gamma} + g^{B\to V}_-(k^2)g_{\alpha\beta}k^{\gamma} + 
g^{B\to V}_0(k^2)p_{\alpha}p^{\beta}q^{\gamma}
\right]. 
\end{eqnarray*}
For the calculation of the $B\to V$ transition form factor $g^{B\to V}_+(0)$, we make use of the same dispersion approach 
of \cite{ma,mb}. The e.m. leptonic decay constant of a vector meson is given by 
\begin{eqnarray}
\langle 0|j^{\rm e.m}_\mu |V(\varepsilon, p)\rangle=\varepsilon_\mu M_V f_V^{\rm e.m.}.  
\end{eqnarray}


\section{Numerical results\label{sec_numerics}}
\subsection{Calculation of the transition form factors}
\subsubsection{Parameters of the model}
The wave function $\phi_B(s)$, can be written as  
\begin{eqnarray}
\label{phi}
\phi_B(s)=
\frac{\pi}{\sqrt2}
\frac{\sqrt{s^2-(m_1^2-m_2^2)^2}}{\sqrt{s-(m_1-m_2)^2}}
\frac{w(k^2)}{s^{3/4}}, \qquad 
k^2=\lambda(s,m_1^2,m_2^2)/4s, 
\end{eqnarray}
with $w(k^2)$ normalized as follows 
\begin{eqnarray}
\label{norm}
\int w^2(k^2)k^2 dk=1. 
\end{eqnarray}
The meson weak transition form factors from dispersion approach reproduce correctly the structure of the heavy quark expansion 
in QCD for heavy-to-heavy and heavy-to-light meson transitions, as well as for the meson-photon transitions, if the 
radial wave functions $w(k^2)$ are localized in a region of the order of the confinement scale, $k^2\le\Lambda^2$ \cite{ma,mb}. 

Following \cite{msa,ms}, we make use of a simple Gaussian parameterization of the radial wave function
\begin{equation}
\label{gauss}
w(k^2)\propto\exp(-k^2/2\beta^2),
\end{equation}
which satisfies the localization requirement for $\beta\simeq \Lambda_{QCD}$ and 
proved to provide a reliable picture of a large family of the transition form factors \cite{ms}. 

In \cite{ms} we fixed the parameters of the quark model -- constituent quark masses and the wave-function parameters $\beta_i$ 
of the Gaussian wave functions -- by requiring that the dispersion approach reproduces  
(i) decay constants of pseudoscalar mesons and (ii) some of the well-measured lattice QCD results for the form factors at large $q^2$. 
The analysis of \cite{ms} demonstrated that a simple Gaussian Ansatz for the radial wave functions allows one to reach this goal 
(to great extent due to the fact that the dispersion representations satisfy rigorous constrains from non-perturbative QCD in the 
heavy-quark limit). With these few model parameters, \cite{ms} gave predictions for a great number of weak-transition form factors 
in the full kinematical $q^2$-region of weak decays. 
However, the analysis of \cite{ms} made use of some approximations which need to be updated: namely, the wave-function parameters 
of $\rho^0$ and $\omega$ mesons, $\beta_\rho$ and $\beta_\omega$, were assumed to be equal to each other, 
and $\beta_\phi$ has been set equal to that of $\eta$ meson. 

In this paper, we make use of the same effective constituent quark masses as obtained in \cite{ms}\footnote{We would like to underline 
that the effective constituent quark masses are used only in the context of the form factor calculations; in the effective $b\to (d,s)$ 
Hamiltonian and for the description of the charm-quark loops, we use the scale-dependent quark masses in the $\overline{\rm MS}$ scheme.} 
\begin{equation}
\label{quark_masses}
\quad m_d=m_u=0.23 \;{\rm GeV}, \quad m_s=0.35 \;{\rm GeV},\quad m_c=1.45 \;{\rm GeV},\quad m_b=4.85 \;{\rm GeV}, 
\end{equation} 
but make the following updates on the determination of the meson wave-function parameters: 
First, we make use of the recent results $f_B=189(4)$ MeV and $f_{B_s}=225(4)$ from lattice QCD \cite{lattice}
to fix the wave-function parameters of $B$ and $B_s$; this new inputs lead to a more reliable 
calculation of the form factors $F_i(q^2,0)$ which involve only $B$-meson wave functions. 
Second, for the calculation of the form factors $F_i(0,q^2)$ based of vector-meson dominance, we  
need the transition form factor $g_+(0)$ for the transition of $B_{s,d}$ meson to neutral vector 
mesons $\rho^0$, $\omega$, $\phi$. 
As mentioned, in \cite{ms} the wave-function parameters of light vector mesons have been set equal to 
the wave-function parameters of 
the corresponding pseudoscalar mesons. This turns out to be a rather crude approximation; 
we now improve the analysis and fix also the vector-meson wave-function parameter $\beta$ from the reproduction of its decay constant. 
The new parameters turn out to be in some cases rather different from those used in \cite{ms}. 

\subsubsection{Form factors $F_i(q^2,0)$}
For the calculation of the form factors $F_i(q^2,0)$, we need the $B_{(s)}$ wave-function parameters, 
which we fix by the requirement to reproduce the latest results for the leptonic decay 
constants $f_B$ and $f_{B_s}$ \cite{lattice}. 
The obtained wave-function parameters and the corresponding decay constants of beauty mesons are quoted 
in Table \ref{table:Bparameters}. 

\begin{table}[H]
\caption{\label{table:Bparameters}
Wave function parameters of $B$ [$M_B=5.279$ GeV] and $B_s$ [$M_{B_s}=5.370$ GeV] mesons and the corresponding 
decay constants calculated with Eq.~(\ref{fP}). For the adopted ranges of 
$\beta$, the corresponding value of the decay constants are shown.}
\centering
\begin{tabular}{|c|c|c|c|c|c|}
\hline
                      & $B$                       &  $B_s$       \\
\hline
$\beta_P$, GeV &  $0.545$-$0.565$ &   $0.61$-$0.63$      \\
\hline
$f_P$, MeV     &  $184$-$192$     &   $221$-$229$       \\
\hline
\end{tabular}
\end{table}

\begin{figure}[ht]
\begin{center}
\begin{tabular}{cc}
\includegraphics[width=8cm,clip]{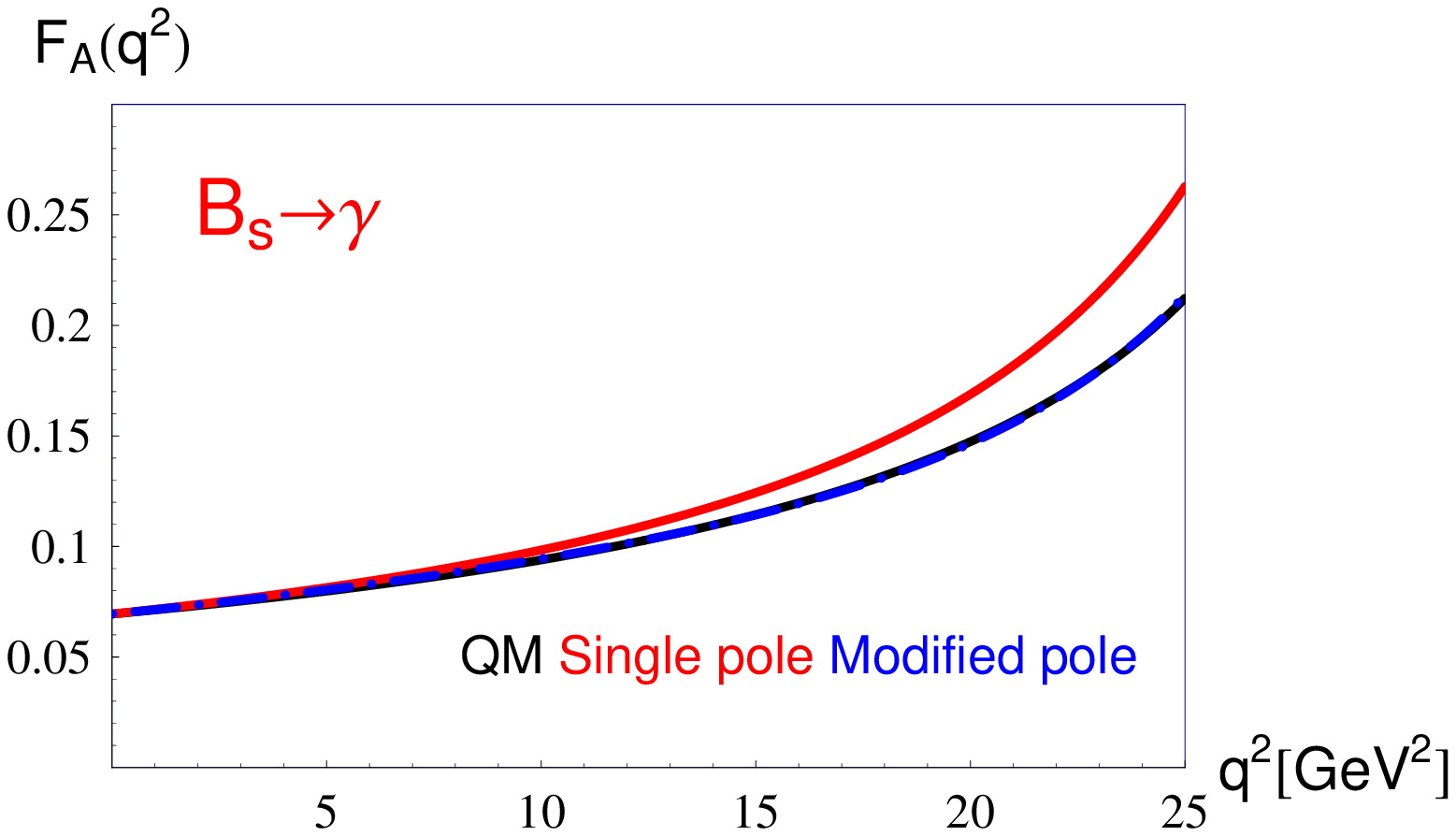} & 
\includegraphics[width=8cm,clip]{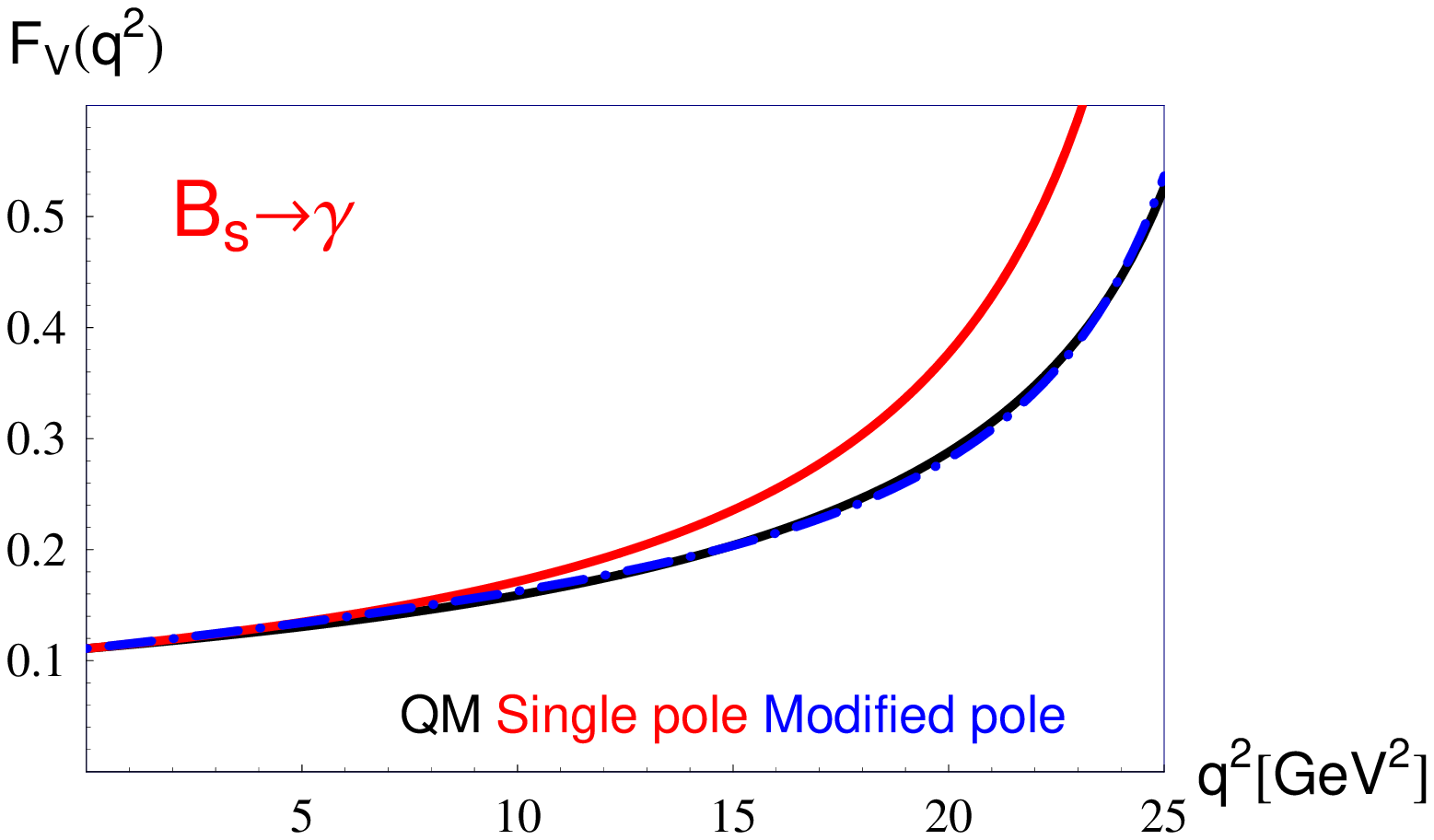}\\
\includegraphics[width=8cm,clip]{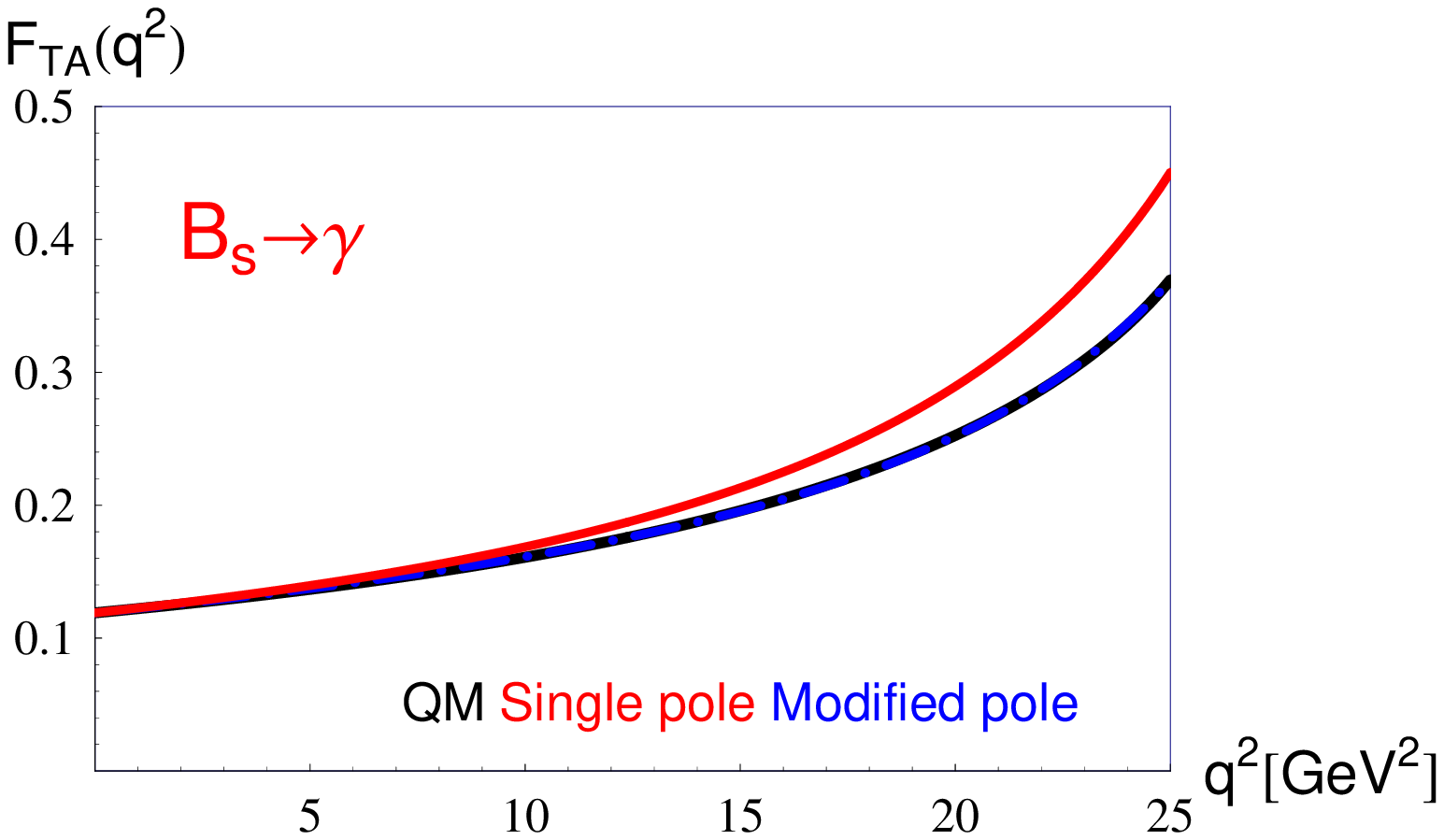} & 
\includegraphics[width=8cm,clip]{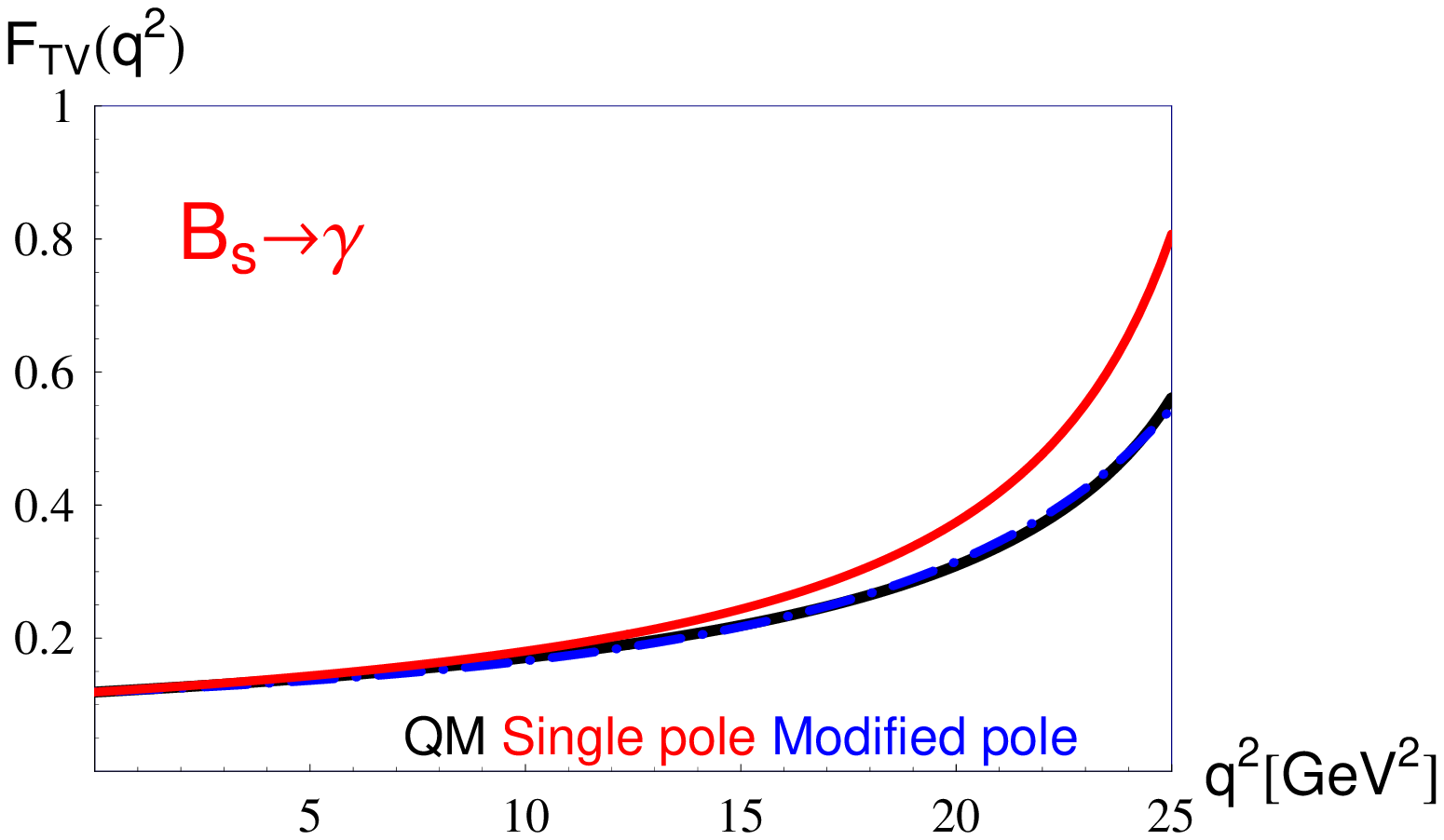}
\end{tabular}
\caption{\label{Plot:ffs}
The form factors $F_i(q^2,0)$ for $B_s\to \gamma$ transitions:
solid black line - the result of the calculation via the dispersion representation; 
blue line - fit to the calculation results with a modified pole formula Eq.~(\ref{modifiedpole}); 
red - a single-pole parametrization.}
\label{Plot:FFS} 
\end{center}
\end{figure}

With the wave function of $B(B_s)$ meson fixed, we calculate the form factors $F^{(b,d)}_i(q^2,0)$ 
($F^{(b,s)}_i(q^2,0)$) via the spectral representations given in the previous Section. 
Fig.~\ref{Plot:ffs} shows the results of our calculation. 

Formally, the spectral representations allow one to calculate the form factors in the region $q^2<(m_b+m_q)^2$. 
However, the results from an approach based on quark degrees of freedom should not be trusted in the region close to hadron 
thresholds. Therefore, one cannot guarantee our calculations to be reliable at $q^2\ge 20$ GeV$^2$. 
On the other hand, we know that the form factors have poles at $q^2=M_R^2$, where $M_R$ is the meson with 
the appropriate quantum numbers: $1^-$ for $F_{V}$ and $F_{TV}$; $1^+$ for $F_{A}$ and $F_{TA}$. 
Therefore, following \cite{ms} we parameterize our results by the ``modified'' pole function 
\begin{equation}
\label{modifiedpole}
F_i(q^2)=\frac{F_i(0)}{(1-{q^2}/{M_{R_i}^2})(1-\sigma_1({q^2}/{M_{R_i}^2})+\sigma_2({q^2}/{M_{R_i}^2})^2)},
\end{equation}
which explicitly takes into account the presence of the pole in the right location. 
Eq.~(\ref{modifiedpole}) approximates the calculation results with better than 3\% accuracy in a broad range 
$0< q^2 < 25$ $GeV^2$. Table \ref{Table:Ffs} gives the corresponding parameters for the $B_{d,s}\to\gamma$ form factors.
We also consider a single-pole parametrization of the form factors 
\begin{equation}
\label{simplepole}
F_i(q^2)=\frac{F_i(0)}{1-{q^2}/{M_{R_i}^2}},
\end{equation}
suggested by large-energy effective theory and used in \cite{Kruger:2002gf}.
At $q^2\le 15$ GeV$^2$ both parametrizations agree with better than 10\% accuracy, which is expected to be the typical systematic 
uncertainty of the form factors from the dispersion approach. At larger $q^2$, the deviation between the two parametrizations
increases, reaching $\sim$ 20\% at $q^2\sim 25$ GeV$^2$. We take this difference as a magnitude of the systematic uncertainty 
of our predictions at large $q^2$.  

\begin{table}[t]
\caption{\label{Table:Ffs}
Parameters of the interpolating formula (\ref{modifiedpole}) for the form factors.}
\centering
\begin{tabular}{|c|cccc|cccc|}
\hline
         &    \multicolumn{4}{c|}{$B_d\to\gamma$}         & \multicolumn{4}{c|}{$B_s\to\gamma$} \\
\hline
         & \hspace{.3cm} $F(0)$ \hspace{.3cm} &  \hspace{.3cm} $\sigma_1$ \hspace{.3cm}  &  \hspace{.3cm}  $\sigma_2$  \hspace{.3cm}&  \hspace{.3cm}  $M_R$[GeV]   
         & \hspace{.3cm} $F(0)$ \hspace{.3cm} &  \hspace{.3cm} $\sigma_1$ \hspace{.3cm}  &  \hspace{.3cm}  $\sigma_2$  \hspace{.3cm}&  \hspace{.3cm} $M_R$[GeV] \\
\hline
$F_A$    &  0.072  &   0.002      &  0.400      &    5.726      &  0.069  & -0.031       &   0.384  &  5.829  \\
$F_V$    &  0.110  &   0.058      &  0.489      &    5.325      &  0.111  &  0.144       &   0.722  &  5.415  \\
$F_{TA}$  &  0.117  &  -0.091      &  0.180      &    5.726      &  0.119  & -0.063       &   0.321  &  5.829  \\
$F_{TV}$  &  0.117  &   0.058      &  0.458      &    5.325      &  0.119  &  0.163       &   0.751  &  5.415  \\
\hline
\end{tabular}
\end{table}
We would like to comment on the systematic uncertainty of our predictions: the constituent quark picture for the form factors 
is an approximation to a very complicated picture in QCD. It is therefore not possible 
to provide any rigorous estimate of the systematic uncertainties of the obtained form factors. 
Based on the comparison of the results from our dispersion approach to the results from QCD sum rules and lattice QCD 
where such results are available, we expect the systematic uncertainty of our form factors at the level of 10\% in the region 
$0< q^2<15$ GeV$^2$ and about 20\% at $15 < q^2$ GeV$^2$.

\subsubsection{Form factors $F_{TA,TV}(0,q^2)$}

For the form factor $F_{TA}(0,q^2)=F_{TV}(0,q^2)$, we make use of the VMD formula (\ref{vmd}) which involves the $B\to V$ transition 
form factor and the decay constants $f_V$. According to adopted procedure, we fix the wave-function parameter $\beta$ for 
neutral vector mesons based on their 
leptonic decay constants, and then, having at hand the wave functions of the $B_{s,d}$ mesons, calculate the $B\to V$ 
transition form factors of interest. 

As inputs for the decay constants of the vector mesons, we make use of two sets of values: 
one set of $f_V$ is obtained directly from the 
experimental data \cite{pdg} neglecting the mixing effects; the second set makes use of the values obtained in \cite{ball2007} 
from the experimantal data including also mixing effects. 
Table~\ref{Table:Vparameters} gives the wave function parameters of neutral vector mesons corresponding to these 
two sets of the decay constants. 

\begin{table}[H]
\caption{\label{Table:Vparameters}
Values of the mass and the width of neutral vector mesons \cite{pdg}, 
decay constant $f_V$ (with isotopic factors omitted) extracted from the leptonic width 
neglecting the meson-mixing effects and the results of \cite{ball2007} which take mixing into account, 
and the corresponding wave function parameters $\beta_V$.}
\centering
\begin{tabular}{|c|c|c|c|c|c|}
\hline
V              &    $M_V$, MeV    & $\Gamma_V$, MeV  &    $f_V$, MeV                  &   $\beta_V$, GeV  \\
\hline       
$\rho^0$       &   775            &         149      &  220 \cite{pdg}, 222 \cite{ball2007}   &   0.328-0.330 \\
$\omega$       &   783            &         8.49     &  195 \cite{pdg}, 187 \cite{ball2007}   &   0.28-0.29  \\
$\phi$         &  1019            &         4.27     &  226 \cite{pdg} 215 \cite{ball2007}    &   0.32-0.34  \\
\hline
\end{tabular}
\end{table}
For the parameters $\beta_V$ we make use of the range, the boundary values of which yield the decay constants 
with/without mixing effects. 
The electromagnetic decay constants which enter the VMD formula (\ref{vmd}) are related to $f_V$ as follows: 
$f^{\rm e.m.}_{\rho^0}=\frac{1}{\sqrt2}f_{\rho^0}$, 
$f^{\rm e.m.}_{\omega}=\frac{1}{3\sqrt2}f_{\omega}$, 
$f^{\rm e.m.}_{\phi}=-\frac{1}{3}f_{\phi}$. 

Having fixed the wave-function parameters, we calculate the transition form factor $g_+$ using the double spectral representation 
for the $P\to V$ form factors given in \cite{melikhov}. The obtained results are summarized in Table~\ref{Table:T1} and compared with the 
determinations from other approaches. We point out a sizeable reduction of the $B_s\to\phi$ form factor compared to the result of \cite{ms}. 
This change just reflects the natural sensitivity of the form factor to the shape of the wave functions of the participating mesons. 
As already pointed out, in \cite{ms} $\beta_\phi$ was assumed equal to $\beta_\eta$, which turns out a rather crude approximation. 
Fixing $\beta_\phi$ from the known value of $f_\phi$, as done in this work, is obviously much more reliable. 
As seen from Table \ref{Table:T1}, our present form-factor results, corresponding to the wave-function parameters 
fixed by using the decay constants as inputs, demonstrate the agreement with the latest results from light-cone sum 
rules, within the expected 10-15\% uncertainty. 
 
We take into account the contributions to the form factor $F_{TA,TV}(0,q^2)$ from the light ground-state vector mesons, 
neglecting those from the excited mesons. The latter are difficult to estimate directly, but the experience from the pion elastic 
and weak transition form factors in the timelike region suggests that the magnitude of the contributions of the excited states is 
numerically much smaller compared to the ground-state contributions \cite{Melikhov:2003hs}. Then, the contributions of the 
excited vector mesons to $F_{TA,TV}(0,q^2)$ may be neglected compared with other contributions to the $B\to\gamma l^+l^-$ amplitude.

\begin{table}[h]
\caption{\label{Table:T1}
The form factor $T_1(0)=-g_+(0)$ from different approaches. The results without the isotopic factors are presented. 
To obtain the form factor for $B\to \rho^0$ one should multiply the entry in the Table by $-1/\sqrt{2}$;
for $B\to \omega$ one should multiply by $1/\sqrt{2}$.}
\centering
\begin{tabular}{|c|c|c|c|c|c|}
\hline
 $T_1(0)$       & This work           & \cite{ms}   & \cite{ball1998}   &  \cite{ball2005} & \cite{ball2007} \\
\hline
$B^+\to\rho^+$ &  $0.29 \pm 0.01$    &  0.27      &  0.29      &  $0.267\pm0.021$ &   0.27(4)        \\
$B\to\omega$  &   $0.24\pm 0.01$     &  $-$       & $ -$       & $0.242\pm 0.022$ &   0.25(4)        \\
$B_s\to\phi$   &   $0.27\pm 0.01$     &  0.38      &  0.35      & $0.349\pm 0.033$ &   0.31(4)       \\
\hline
\end{tabular}
\end{table}

%
%
%
%
\subsection{\label{sec4.B}The differential distributions}

With all form factors known, we are in a position to calculate numerous differential distributions in $B\to \gamma l^+l-$ decays.
The necessary inputs such as the $V_{\rm CKM}$ and the quark masses are taken from \cite{pdg}; the values of 
the Wilson coefficient are summarized at the end of Section 2. The only theoretical ingredient which still contains ambiguities 
is the contribution of charm-loops in the charmonia resonance region. 
As discussed above in Section \ref{sec:cc}, one can obtain an excellent description of the QCD sum rule results \cite{hidr} 
available at low $q^2$ with different assignments of the resonance phases. The phase ambiguity cannot be resolved on the basis 
of the theoretical arguments only, but needs further inputs from the experimental measurements which will become available in the future. 
Most of the differential distributions presented below are obtained for the standard assignment of positive contributions 
of all charmonia, that follows the patters of the factorizable contributions. 
The only exception is the forward-backward asymmetry, $A_{\rm FB}$, in which case we discuss two different assignments and demonstrate a strong 
sensitivity of $A_{\rm FB}$ in the $q^2$-region between $\psi$ and $\psi'$ to the specific choice of the resonance relative signs. 

\subsubsection{\label{sec4.1}The differential branching ratios}

\begin{figure}[H]
\begin{center}
\begin{tabular}{cc}
\mbox{\centering
\includegraphics[width=8cm]{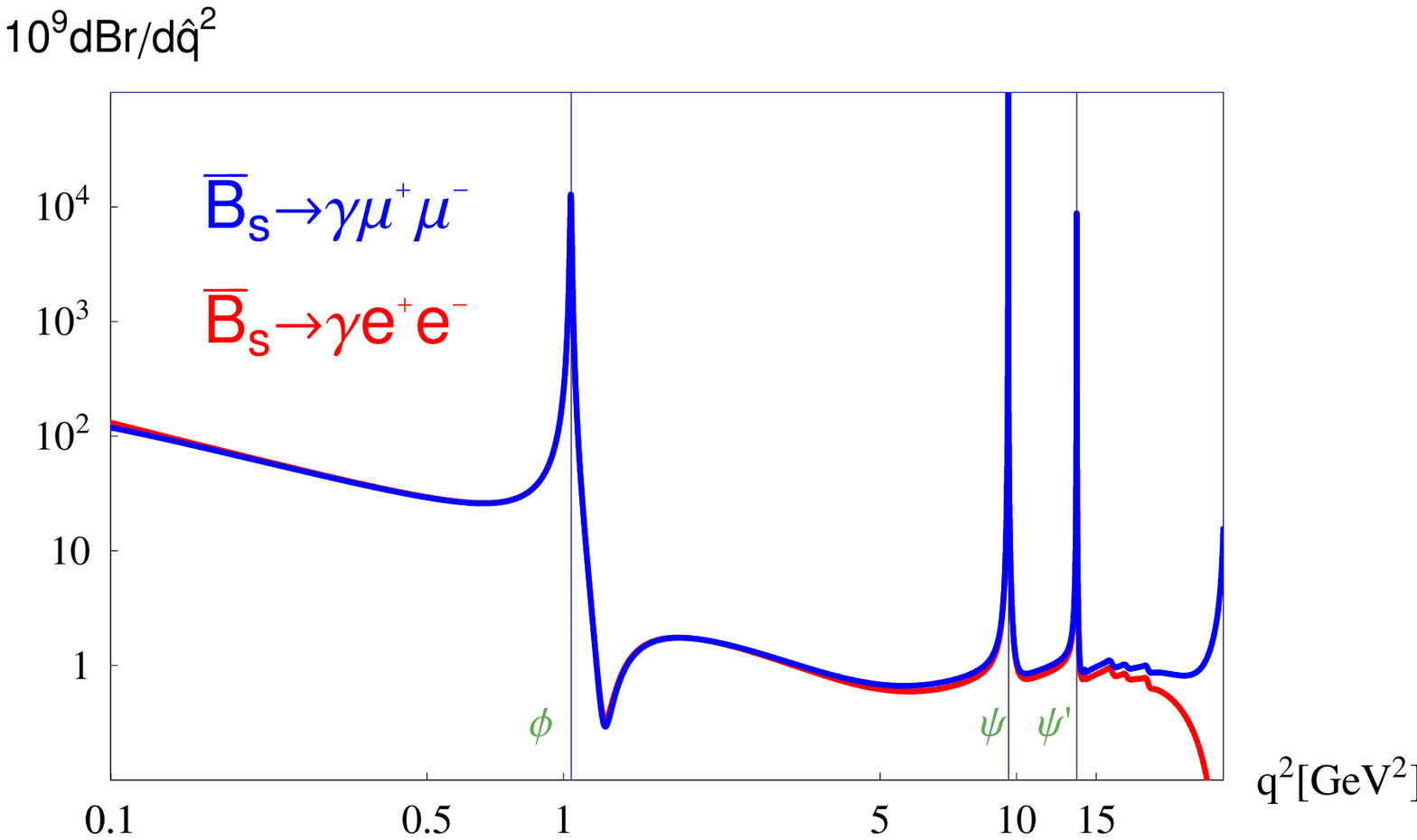}} &
\mbox{\centering
\includegraphics[width=8cm]{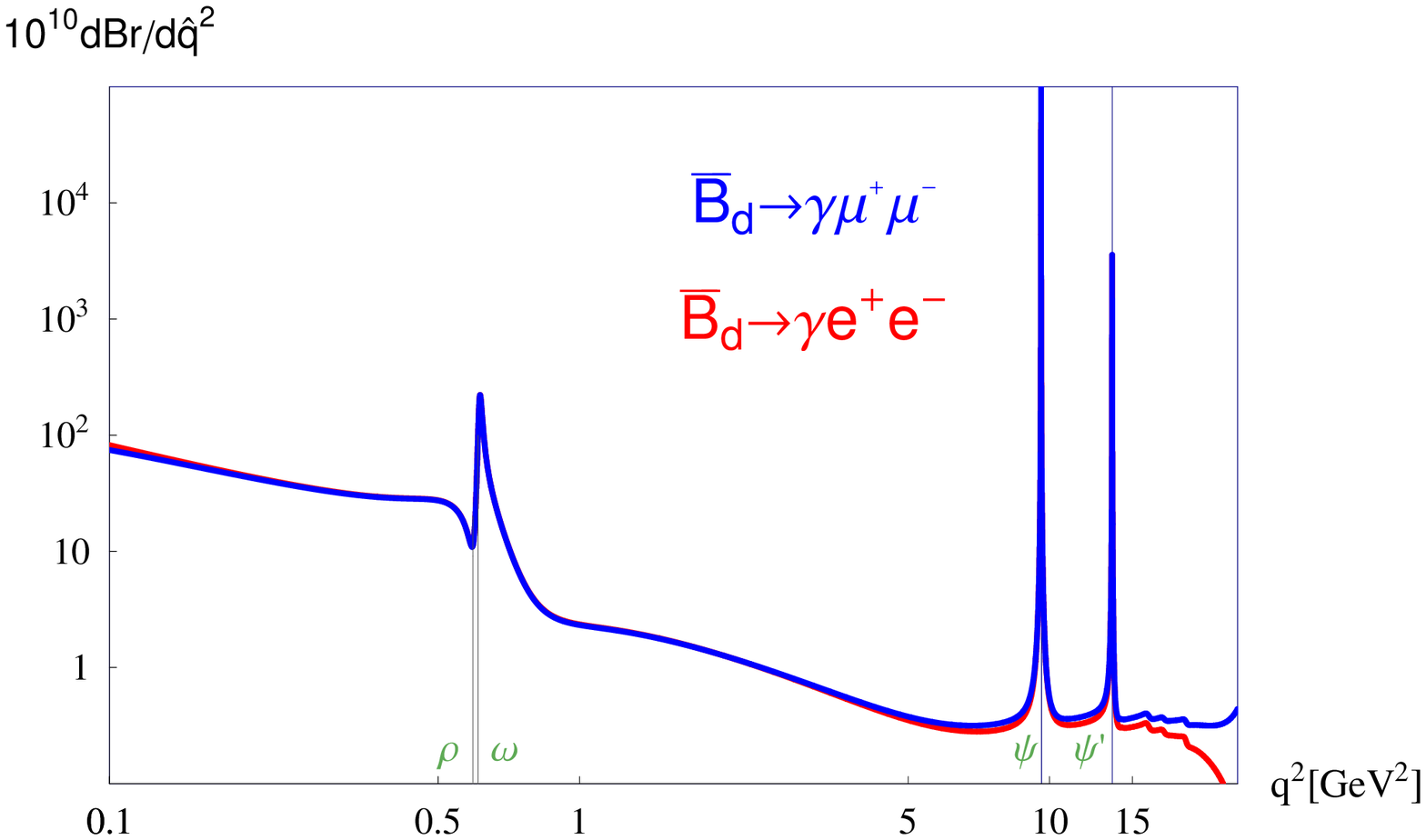}} 
\end{tabular}
\caption{Differential branching fractions for $B_s\to \gamma l^+l^-$ (left) and $B_d \to \gamma l^+l^-$ (right) decays.
Blue lines - $\mu^+\mu^-$ final state, red lines - $e^+e^-$ final state.}
\label{Plot:BR} 
\end{center}
\end{figure}

The differential branching ratios are shown in Fig.~\ref{Plot:BR}. The results in Fig.~\ref{Plot:BR} 
correspond to the description of the charm-loop effects according to Eq.~(\ref{Hdisp}) and adding the contributions of the broad charmonia 
according to Eq.~(\ref{h}), and further assuming that all charmonia contribute with the same positive sign 
(coinciding with the sign of the factorizable contribution). The subtraction constants $a$ and $b$ in Eq.~(\ref{Hdisp}) are determined 
by the requirement to reproduce the known results at $q^2\le 4m_c^2$, including nonfactorizable corrections calculated in \cite{hidr}.
As discussed in Section \ref{sec:cc}, this reproduction 
may be reached with an excellent few percent accuracy for the different assignments of the resonance phases thus leaving the question 
of the relative resonance phases open. 

In the region $q^2\le 6$ GeV$^2$, the charming loops provide a mild contribution at the level of a few percent, and therefore 
the branching fractions in this region may be predicted with a controlled accuracy, mainly limited by the form-factor uncertainty.
Our estimates read
\begin{eqnarray}
\label{Br1-6GeV2}
&&{\cal B}(\bar B_s\to \gamma l^+l^-)|_{q^2\in[1,6]\, {\rm GeV}^2}=(6.01\pm 0.08\pm 0.70)10^{-9}\nonumber\\
&&{\cal B}(\bar B_d\to \gamma l^+l^-)|_{q^2\in[1,6]\, {\rm GeV}^2}=(1.02\pm 0.15\pm 0.05)10^{-11}.
\end{eqnarray}
The first uncertainty in these estimates reflects merely the 10\% uncertainty in the $B\to\gamma$ transition form factors. 
The second uncertainty reflects the uncertainty in the contributions of the light vector mesons $\rho$, $\omega$, and $\phi$. 
We would like to emphasize that in the case of the $B_s\to \gamma l^+l^-$ transition, the dominant contribution is 
given by the narrow $\phi$-meson pole. In the case of the $B_d\to \gamma l^+l^-$ transition, the known contribution of 
the vector resonances is less important, and as the result, the branching ratio uncertainty reflects to a large extent 
the form factor uncertainty of 10\%. 

Table~\ref{tab-5a} gives the branching ratios integrated over several bins below the charmonia region (the charmonia region 
$0.33 \le q^2/M_{B_s}^2 \le 0.55$ is normally excluded from the data analysis). Table~\ref{tab-5b} shows the results 
of the integration above the charmonia region, from $q^2=0.55 M_{B_s}^2$ to $q^2=M_B^2-2M_B E^{\gamma}$ for different values of $E^{\gamma}$, 
the photon energy in the $B$-meson rest frame: since the Bremsstrahlung contribution leads to a divergence of the branching ratio 
at large $q^2$, a certain cut on the photon energy is required. The values in the range $0.1\,\textrm{GeV}< E^\gamma_{min} < 0.5\,\textrm{GeV}$ 
correspond to the photon selection criteria at the Belle II detector \cite{Aushev:2010bq}, 
while the interval  $0.5\,\textrm{GeV}< E^\gamma_{min} < 1\,\textrm{GeV}$ is relevant for those at 
the LHCb detector \cite{LHCB:detector,LHCB:detectorb}. 
\begin{table}[!t]
\centering
\caption{\label{tab-5a}
The contributions to the branching ratio of $\bar B_{s}\to\gamma l^+l^-$ decays integrated over the specific $q^2$-ranges
in the region below charmonia resonances for the central values of all parameters and the form factors given by Eq.~(\ref{modifiedpole}).}
\begin{tabular}{|r|c|c|c|c|}
\hline
                      &  \multicolumn{4}{c|}  { $10^9\Delta {\cal B}(B_s\to \gamma l^+l^-)$}  \\
\hline
                      & $[4m_e^2,4m_\mu^2]$ &  $[4m_\mu^2,1\,{\rm GeV}^2]$ & $[1\,{\rm GeV}^2,6\,{\rm GeV}^2]$  &  $[6\,{\rm GeV}^2,0.33 M^2_{B_s}]$ \\
\hline
$e^+e^-$      &  4.672            &   1.796                   &  6.003                             &       0.136         \\
$\mu^+\mu^-$  &  $-$              &   1.790                   &  6.004                             &       0.149         \\
\hline
\end{tabular}
\centering
\caption{\label{tab-5b}
The contributions to the branching ratio of $B_{s}\to\gamma l^+l^-$ decays from the $q^2$-region $[0.55 M^2_{B_s}, M^2_{B_s}-2M_{B_s}E_\gamma]$.}
\begin{tabular}{|r|c|c|c|c|}
\hline
                      &  \multicolumn{4}{c|}  {$10^9\Delta {\cal B}(B_s\to \gamma l^+l^-)$}  \\
\hline
  $E_\gamma$[GeV] & 0.08 &  0.1  & 0.5   &  1.0   \\
\hline
$e^+e^-$ &\hspace{.2cm} 0.20 \hspace{.2cm} &\hspace{.2cm} 0.20\hspace{.2cm} &\hspace{.2cm} 0.16 \hspace{.2cm} &  \hspace{.2cm} 0.06  \hspace{.2cm} \\
$\mu^+\mu^-$  &  0.43     &   0.41   & 0.23 &       0.07         \\
\hline
\end{tabular}
\end{table}

In the actual data analysis, the photon-energy cut is applied in the laboratory frame, not in the $B$-rest frame. 
However, as can be seen from Tables \ref{tab-5a} and \ref{tab-5b}, 
the specific value of the energy have a marginal impact on the total branching ratio, since the contribution of the region 
above charmonia resonances, after applying the cuts, contributes to the branching ratio at the level below 10\%. 


\subsubsection{\label{sec4.2}Ratio $R_{\mu/e}$ of the muon/electron distributions in $B\to\gamma l^+l^-$ decays}
\begin{figure}[b]
\begin{center}
\begin{tabular}{cc}
\includegraphics[width=6cm]{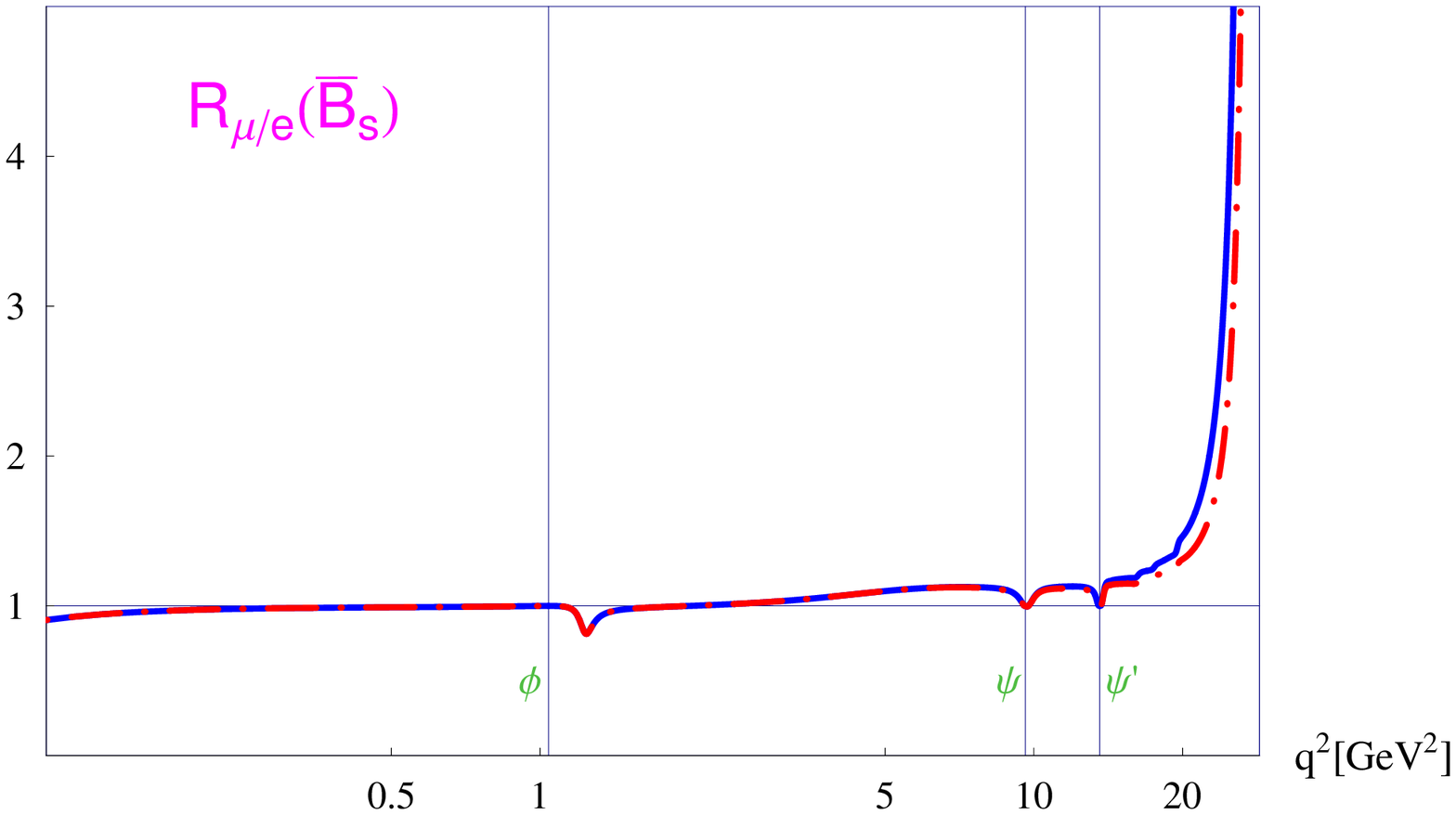} & \includegraphics[width=6cm]{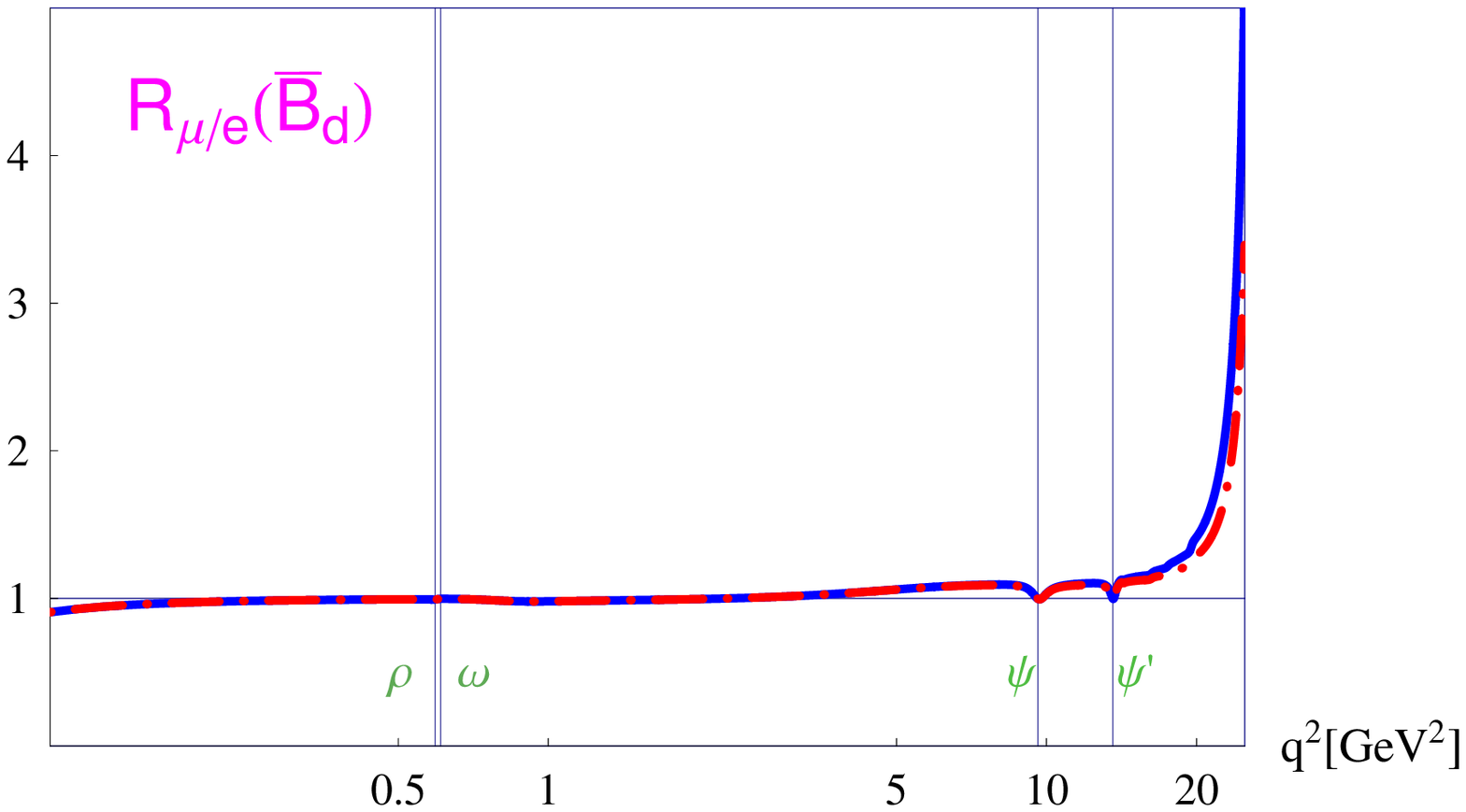}\\
(a)  &  (b)  \\
\includegraphics[width=6cm]{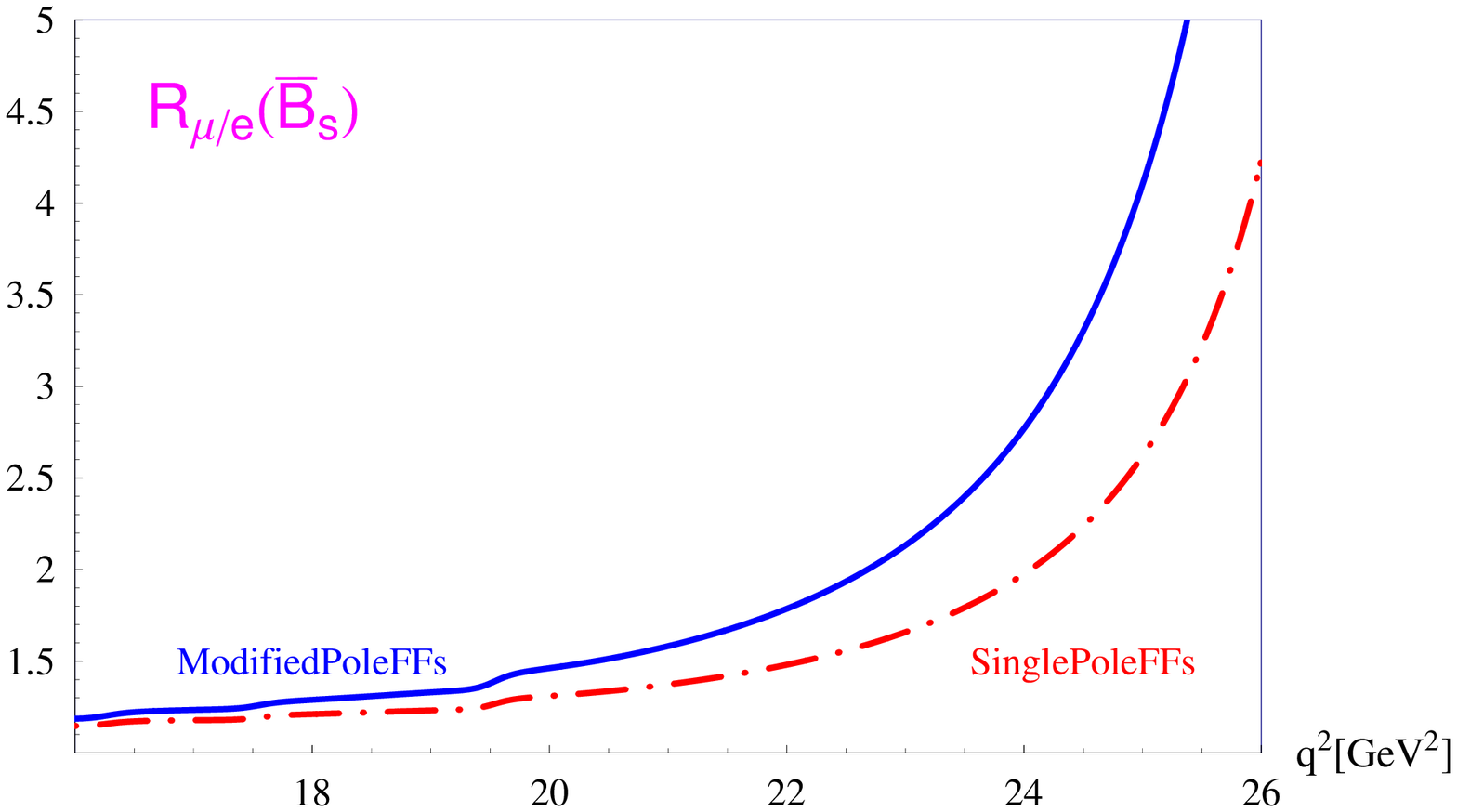} & \includegraphics[width=6cm]{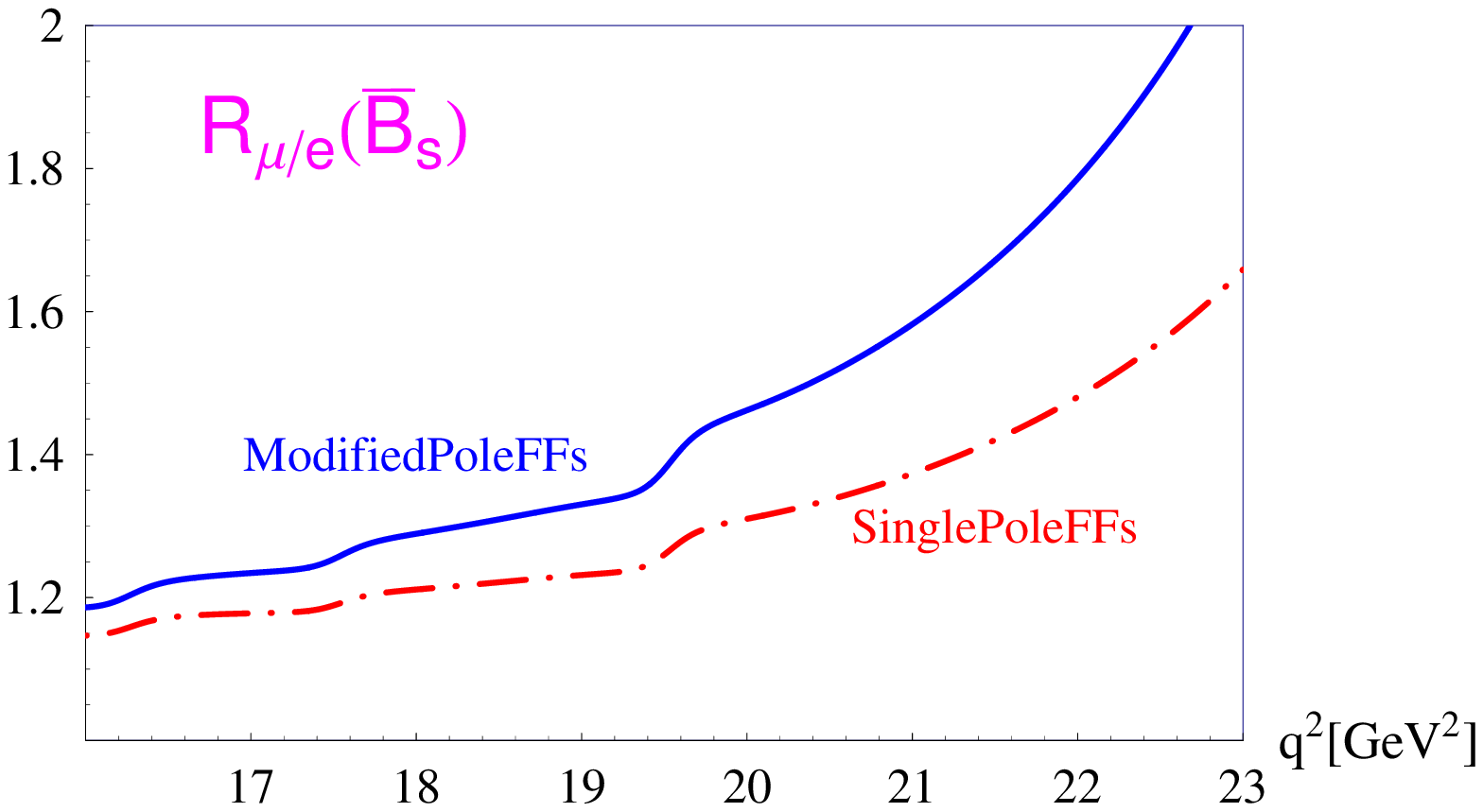}\\
(c)  &  (d)  
\end{tabular}
\caption{The ratio $R_{\mu/e}$ in $\bar B_s\to\gamma l^+l^-$ (a) and $\bar B_s\to\gamma l^+l^-$ decays (b). 
The sensitivity of $R_{\mu/e}(\bar B_s)$ at large $q^2$ to the $q^2$-dependence of the $B_s\to\gamma$ transition form factors (c,d).}
\label{Plot:R} 
\end{center}
\end{figure}

Fig.~\ref{Plot:R} shows the ratio $R_{\mu/e}$ of the differential distributions $\bar B\to \gamma\mu^+\mu^-$ to $\bar B\to \gamma e^+e^-$. 
Such a ratio is a standard observable for probing violations of the lepton universality in rare FCNC semileptonic 
decays $B\to (P,V)l^+l^-$; recently,  
this ratio was proposed as a useful variable also in radiative leptonic decays \cite{gz2017}. 
The ratio is found to be close to unity in the region 
$q^2\le 5$ GeV$^2$. Notice however, that in $B\to\gamma l^+l^-$ decays, unlike the $B\to(P,V)l^+l^-$ processes, 
the lepton-mass effects, mainly the Bremsstrahlung contribution to the amplitude, come into the game. 

At large $q^2$, the terms proportional to $C_{7\gamma}$ can be neglected compared to those containing $C_{9V}$ and $C_{10A}$ and 
one obtains the following expressions for different contribition to the decay rate:

\begin{eqnarray}
&&\Gamma_1=\frac{\alpha_{e.m.}^3G_F^2 |V_{tb}V_{tq}|^2}{768\pi^4}
({C^{\rm eff}_{9V}}^2+C_{10A}^2)(F_A^2+F_V^2)(1-q^2/M_B^2)^3M_B^5,
\nonumber\\
&&\Gamma_{2}=\frac{\alpha_{e.m.}^3G_F^2 |V_{tb}V_{tq}|^2}{8\pi^4}
\frac{\left\{2 \log(M_B/m_l)-1\right\}C_{10A}^2f_B^2m_l^2M_B}{1-q^2/M_B^2},
\nonumber\\
&&\Gamma_{12}=-\frac{\alpha_{e.m.}^3G_F^2 |V_{tb}V_{tq}|^2}{16\pi^4}\log(M_B/m_l)
C^{\rm eff}_{9V}C_{10A}f_B m_l^2 M_B^2 F_V (1-q^2/M_B^2)^2.
\end{eqnarray}
For the electrons in the final state, all terms containing $m_l$ may be neglected; for the muons $\Gamma_{12}$ gives a negligible 
contribution compared to $\Gamma_{1}$ and $\Gamma_{2}$. Finally, for $q^2$ above the narrow charmonia region, one finds an  
approximate relation 
\begin{eqnarray}
R_{\mu/e}=1+
\frac{96\left\{2 \log(M_B/m_\mu)-1\right\}C_{10A}^2}{{C^{\rm eff}_{9V}}^2+C_{10A}^2}
\frac{1}{(F_A^2+F_V^2)(1-q^2/M_B^2)^4}
\frac{m_\mu^2f_B^2}{M_B^4}. 
\end{eqnarray}
Obviously, the Bremsstrahlung contribution drives the ratio $R_{\mu/e}$ far above unity at $q^2\to M_B^2$.
In practice, at $q^2\ge 15$ GeV$^2$ the ratio $R_{\mu/e}$ is sensitive to the details of the $q^2$-behaviour of 
the transition form factors $F_{A,V}$. So, if the lepton universality has been verified at small $q^2$, 
measuring $R_{\mu/e}$ at large $q^2$ provides a direct access to the $q^2$-dependence of the $B\to\gamma$ transition form factors. 

\subsubsection{Forward-backward asymmetry \label{sec4.3}}

The differential forward-backward asymmetry is given by  the relation
\begin{eqnarray}
A_{FB}(\hat{s})\,=\,\frac{\int\limits_0^1d\cos\theta \, \frac{d^2\Gamma(B_{(s)}\to l^+l^-\gamma)}{d\hat{s} \, 
d\cos\theta}-\int\limits_{-1}^0d\cos\theta \, \frac{d^2\Gamma(B_{(s)}\to l^+ l^-\gamma)}{d\hat{s} \, 
d\cos\theta}}{\frac{d\Gamma(B_{(s)}\to l^+l^-\gamma)}{d\hat{s}}},
\end{eqnarray}
where $\hat{s}\,=\,q^2/M_B^2$,\, $\theta$ is the angle between $\vec{p}$ and $\vec{p_2}$, the momentum of the 
negative-charge lepton. 

\begin{figure}[t]
\begin{center}
\begin{tabular}{cc}
\includegraphics[width=8cm,clip]{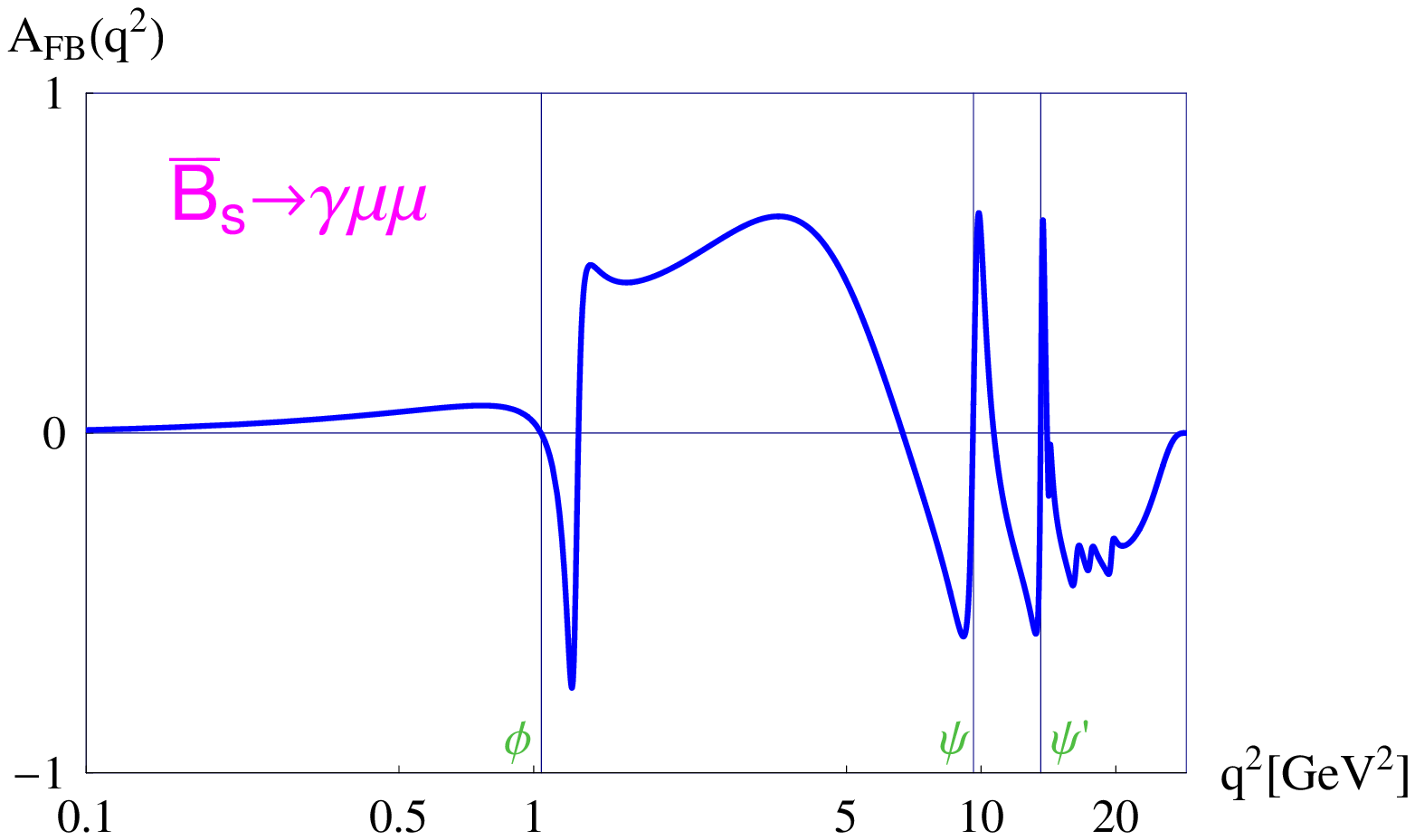} & 
\includegraphics[width=8cm,clip]{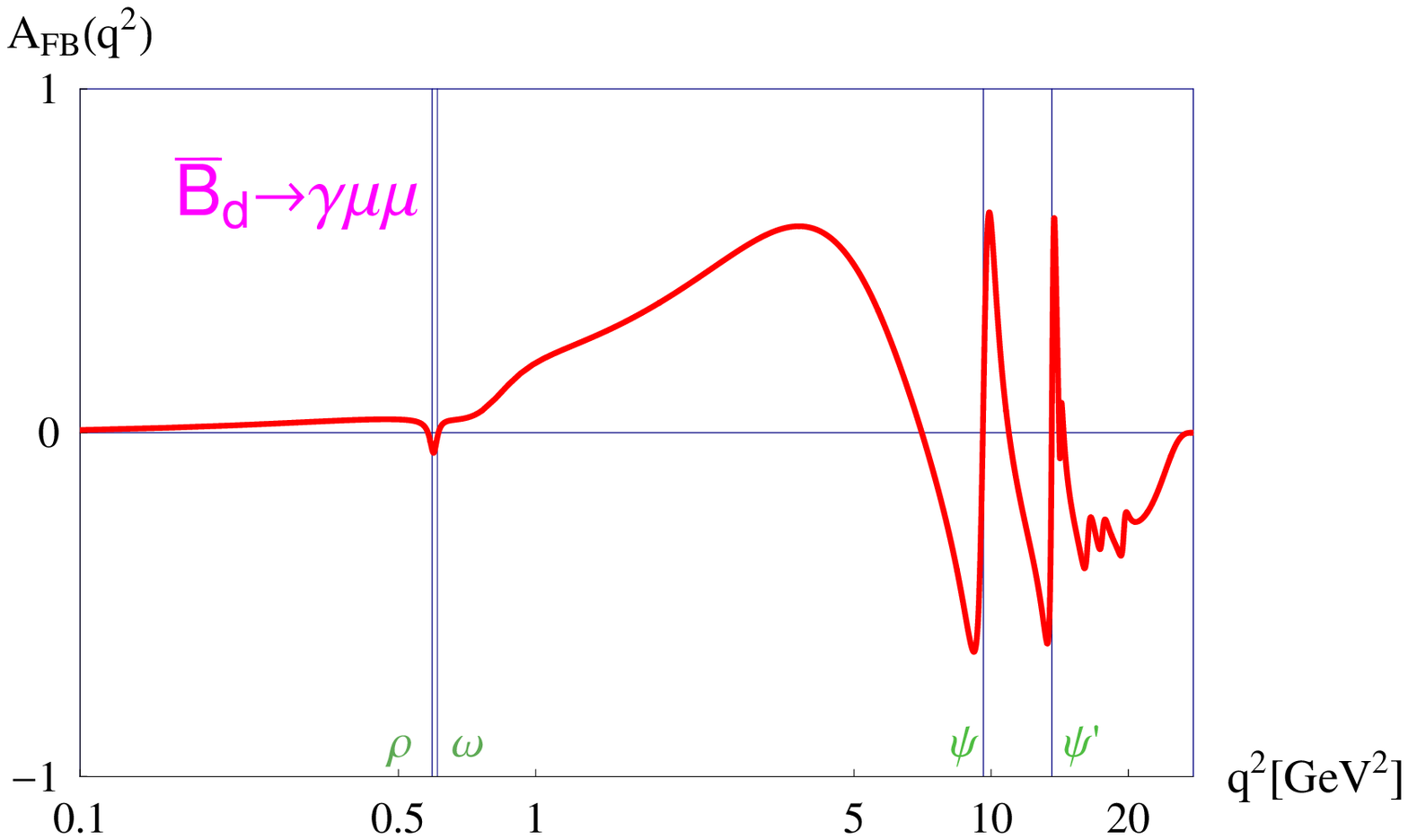}\\
\includegraphics[width=8cm,clip]{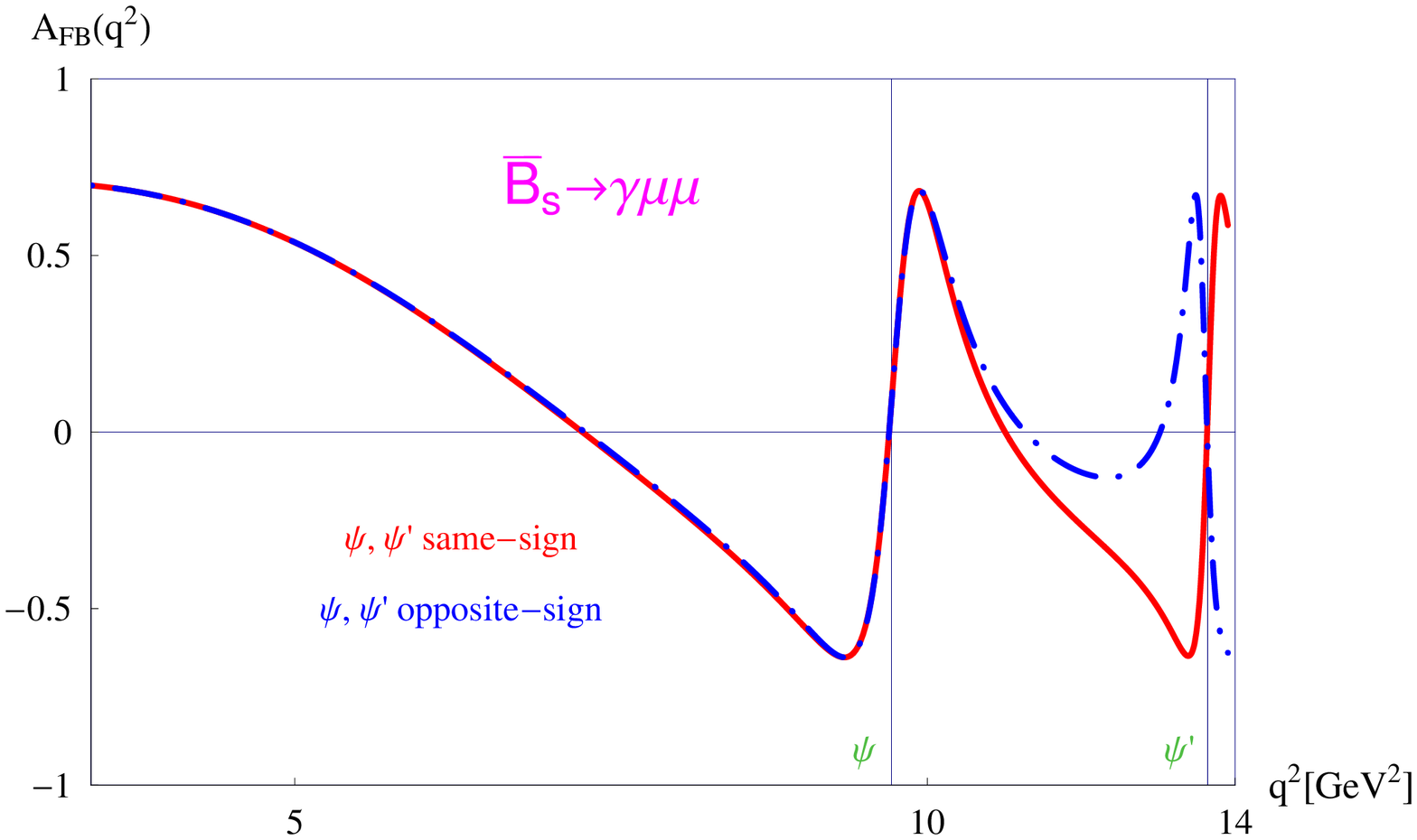} & 
\includegraphics[width=8cm,clip]{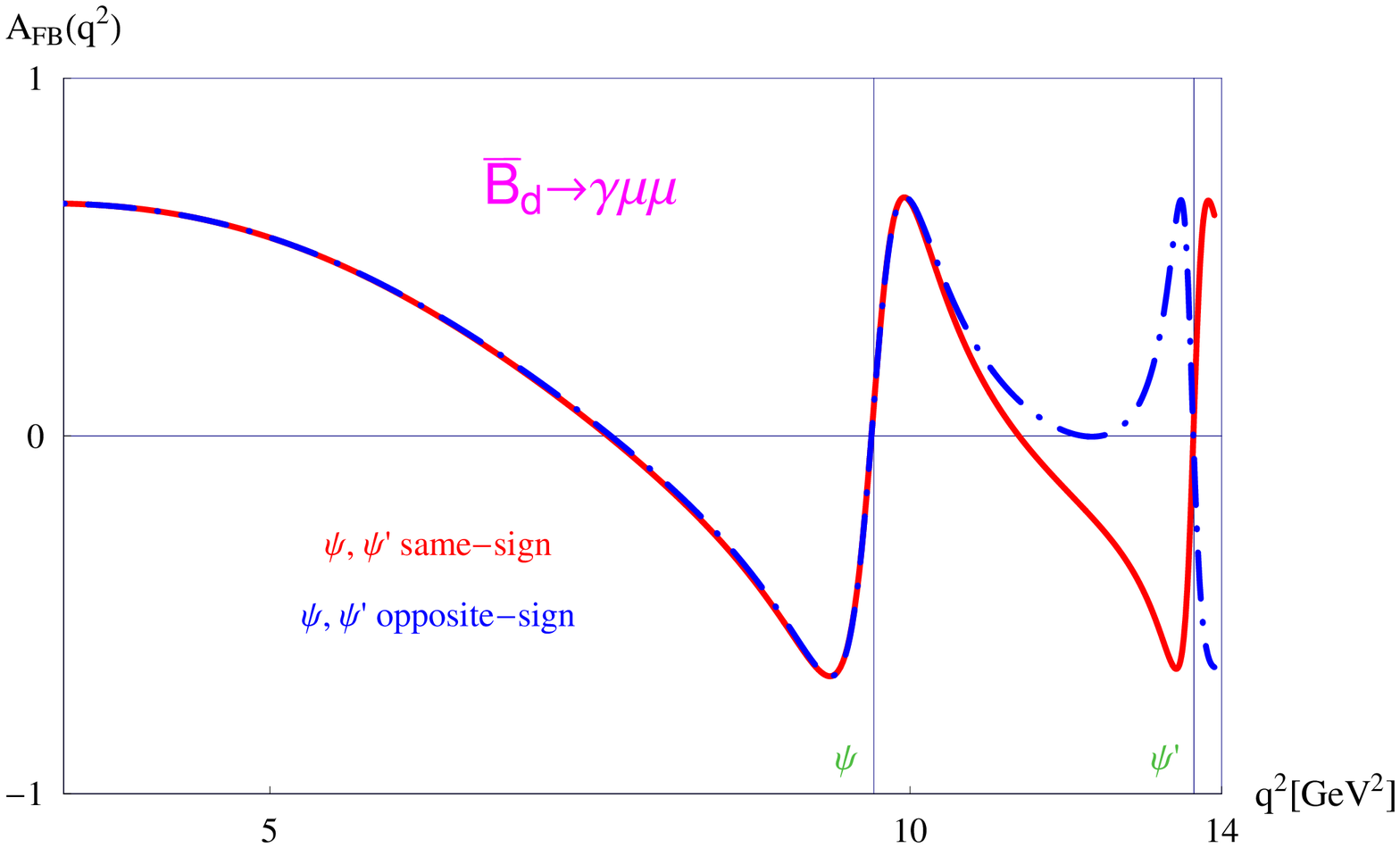}
\end{tabular}
\caption{Forward-backward asymmetry for $\bar B_s\to \gamma \mu^+\mu^-$ (left column) and $\bar B_d\to \gamma \mu^+\mu^-$ (right column) decays.
The lower plots show the asymmetries at $q^2< M^2_{\psi'}$ for two different relatives signs of $\psi$ and $\psi'$ contributions.}
\label{Plot:AFB} 
\end{center}
\end{figure}

The results for $A_{FB}$ in the $\bar B_s\to \gamma\mu\mu$ and in the $\bar B_d\to \gamma\mu\mu$ are shown in 
Fig.~\ref{Plot:AFB} (a) and (b), respectively. The asymmetries are practically insensitive to the uncertainties in the $B\to\gamma$ 
transition form factors, as these uncertainties to large extent cancel each other in the asymmetries. 

As discussed above, the relative phases of the charmonium resonances cannot be unambiguously determined on the basis of the 
QCD sum-rule calculation of the nonfactorizable corrections at $q^2\le 4m_c^2$. The results shown in Fig.~\ref{Plot:AFB} (a) and (b),
are obtained making use of the QCD sum-rule results for nonfactorizable contributions 
at small $q^2$ and employing the conventional assignment of the signs of all charmonia contributions to be positive, 
following the patter of the factorizable contributions. 

It should be emphasized, that $A_{FB}$ in the region between $\psi$ and $\psi'$ provides an unambiguous test of the relative 
signs of $\psi$ and $\psi'$ contributions: as displayed in Fig.~\ref{Plot:AFB} (c) and (d), $A_{FB}$ being insensitive to the patter 
of charmonium resonances in $C_{9V}$ in the region of small $q^2$, 
demonstrates qualitative differences depending on the relative signs between 
$\psi$ and $\psi'$ in the region between the resonances. 
Therefore, experimental study of the asymmetry in this region will allow one to check the relative signs of the $\psi$ and $\psi'$.
We point out that this observation applies not only to $B\to \gamma l^+l^-$ decays, but also to $B\to Vl^+l^-$ decays \cite{cc_2}, 
where such measurement seems much more feasible.

One should take into account, however, that the measurement of the forward-backward asymmetry in the 
$B\to \gamma l^+l^-$ decay 
seems to be a hard 
task, because the final state $\gamma l^+l^-$ does not carry any information about the flavor 
of the decaying $B$-meson. In addition, the signs of the asymmetries corresponding to $B$ and $\bar B$ mesons are opposite. 
In the absence of flavor tagging, the total asymmetry equals zero aside from CP-violating effects. 
It appears that flavor tagging is impossible at LHCb; however, at Belle II one can use the fact that neutral 
$B$-mesons are produced in an entangled state. Thus, if one of the $B$-mesons decays to a state with a certain flavor, 
the other $B$-meson decaying to $\gamma l^+l^-$ has the opposite flavor. Now, if the interval between the 
decays is less than half of the oscillation period, one can claim that the flavor of the second $B$-meson is 
known with sufficient probability. For each selection procedure one can also account for the oscillations 
contribution and therefore improve the prediction accuracy. A method for such calculations was developed in \cite{Balakireva:2009kn}.

\section{Conclusions\label{conclusions}\label{sec5}}

The paper presents a detailed analysis of nonperturbative QCD effects in rare 
FCNC decays $B_{(s,d)}\to \gamma l^+l^-$ decays in the Standard model. 
It should be emphasized that FCNC radiative leptonic decays exhibit a 
much more diverse structure of such effects compared to 
FCNC semileptonic decays. 

\vspace{.2cm}

Our main results may be summarized as follows: 
\vspace{.2cm}

\noindent 
1. We analyzed the $B\to\gamma^*$ transition amplitudes induced by the the vector, 
axial-vector, tensor, and pseudotensor $b\to (s,d)$ quark 
currents. The invariant form factors, which parameterize the corresponding amplitudes, 
depend on two variable $k'^2$ and $k^2$, 
where $k'$ is the momentum emitted from the FCNC $b\to (s,d)$ vertex, and $k$ is the momentum of the virtual photon. 
We worked out the constraints on these form factors imposed by electromagnetic gauge invariance. 
We then calculated all the form factors of interest making use of the relativistic dispersion approach 
based on the constituent quark picture. The appropriate subtractions in the spectral representations for 
the form factors have been determined from the constraints imposed by gauge invariance. 
\vspace{.1cm}

For those form factors, describing the processes with $l^+l^-$ pair emitted from the FCNC vertex, 
the numerical results in the full $q^2$-range of the $B\to\gamma$ decay have been obtained entirely 
within the dispersion approach. The resulting $q^2$-dependences of the form factors 
exhibit the expected properties, in particular, suggest the pole beyond the physical decay 
region at the right location required by the quantum numbers of the appropriate meson resonances. 
\vspace{.1cm}

For those form factors, describing the $l^+l^-$ pair emission from the {\it light} valence quark 
of the $B$-meson, one encounters 
the light vector-meson resonances in the physical decay region. To obtain the predictions for the 
form factors in this case, 
we combined the results from the direct calculation within our dispersion approach with the 
gauge-invariant version of the vector-meson 
dominance model. The numerical predictions for the form factors involving vector mesons have
been updated by using the new procedure 
of fixing the wave-function parameters of vector mesons: namely, the wave-function parameters of the vector mesons involved have 
been fixed by requiring that the decay constants reproduce the latest results for $f_V$ known from other theoretical approaches. 

\vspace{.1cm}
The weak form factors of heavy mesons obtained in the dispersion approach satisfy rigorious contraints known from QCD in the heavy-quark limit. 
Still, the dispersion approach is a phenomenological approach representing a specific formulation of the relativistic quark model. 
Therefore, essentially the only way to probe its accuracy is a comparison of its predictions with the predictions from direct QCD 
related approaches. A comparison with the known results from lattice QCD and QCD sum rules, where such results are available, 
suggest that the uncertainty of the form factors from the dispersion approach does not exceed the level of about 10\%. 
We therefore assign this accuracy to our predictions for the $B\to\gamma$ transition form factors. 
\vspace{.2cm}

\noindent 
2. We performed a detailed study of the charm-quark contributions and 
worked out the general constraints on these contributions imposed by electromagnetic gauge invariance. 
Assuming the similarity of the charm-loop contributions to $B\to\gamma l^+l^-$ and to $B\to V l^+l^-$ amplitudes,  
$V$ the vector meson, we obtained numerical predictions for the charm contributions to the $B\to\gamma l^+l^-$ amplitude making 
use of the existing estimates of the nonfactorizable effects in the $B\to V l^+l^-$ decays. 
We demonstrated that the results for the nonfactorizable corrections available at $q^2$ below the charm threshold do not allow one to resolve 
the possible ambiguity in the relative charmonium resonance phases. Additional inputs are necessary to determine the phases 
unambiguously. 
We have shown that the forward-backward asymmetry in the $q^2$-range between $\psi$ and $\psi'$ provides an efficient probe of the 
relative charmonium contributions. 
\vspace{.2cm}

\noindent 
3. We obtained numerical predictions for a number of the differential distributions in $B\to\gamma l^+l^-$ decays. In particular, we 
demonstrate that $R_{\mu/e}(q^2)$, the ratio of the $B\to\gamma \mu^+\mu^-$ over $B\to\gamma e^+e^-$ differential 
distributions, at large $q^2$ provides direct access to measuring the $q^2$-dependences of the $B\to \gamma$ transition form factors, 
once the lepton universality is established from the data at low $q^2$. 

We also calculated the branching ratios integrated over the low-energy range $q^2=[1,6]$ GeV$^2$ where 
(i) the form factors are known reliably and (ii) the $\bar cc$ contributions remain at the level of a few percent:  
\begin{eqnarray}
&&{\cal B}(\bar B_s\to \gamma l^+l^-)|_{q^2\in[1,6]\, {\rm GeV}^2}=(6.01\pm 0.08\pm 0.70)10^{-9},\nonumber\\
&&{\cal B}(\bar B_d\to \gamma l^+l^-)|_{q^2\in[1,6]\, {\rm GeV}^2}=(1.02\pm 0.15\pm 0.05)10^{-11}.\nonumber
\end{eqnarray}
The first error in these predictions reflects the uncertainty of the $B\to\gamma$ transition form factors; the second error 
reflects the uncertainty in the contributions of the light vector resonances ($\rho,\omega,\phi$). 


\vspace{.5cm}

\noindent {\bf Acknowledgements}
\vspace{.2cm}

\noindent
The authors have pleasure to thank Damir Becirevic, Diego Guadagnoli, Otto Nachtmann, and Berthold Stech 
for interesting and stimulating discussions. We are particularly grateful to Diego Guadagnoli for valuable 
comments on the initial version of this paper. 
D.~M. was supported by the Austrian Science Fund (FWF) under project P29028. 
A.~K. is grateful to the ``Basis'' Foundation (Russia) for a stipendium for PhD students. 
Sections 1-3, 6, 8, and 9 were done with the support of grant 16-12-10280 of the Russian Science Foundation (A.K. and N.N.).
 


\end{document}